\documentclass[twocolumn,amsmath,amssymb,aps,preprintnumbers,prd]{revtex4-1}
\usepackage{graphicx} 
\usepackage{color}
\usepackage{slashed} 
\usepackage{hyperref}
\allowdisplaybreaks
\def\beq{\begin{equation}}
\def\eeq{\end{equation}}
\def\beqn{\begin{eqnarray}} 
\def\eeqn{\end{eqnarray}}

\newcommand{\salamanca}{Departamento de F\'isica Fundamental e IUFFyM, Universidad de Salamanca, Plaza de la Merced S/N, 37008 Salamanca, Spain.}

\begin{document}
\title{Analysis of symmetries in the Causal Loop-Tree Duality representations}

\author{Irene Lopez Imaz$^{(a)}$} \email{ireneimaz@usal.es} 
\author{German F.R. Sborlini~$^{(a)}$} \email{german.sborlini@usal.es} 
\affiliation{$^{(a)}$ \salamanca\\}

\begin{abstract}
Unveiling hidden symmetries within Feynman diagrams is crucial for achieving more efficient computations in high-energy physics. In this paper, we study the symmetries underlying the causal Loop-Tree Duality (LTD) representations through a {graph-theoretic} analysis. Focusing on the integrand-level representations of $N$-point functions at one loop, we examine their degeneration and discover that different causal representations are interconnected through specific transformations arising from the symmetries of cut diagrams. Furthermore, the degeneration is linked to algebraic constraints among the different causal thresholds. Our findings shed new light on the deeper structures of Feynman integrals and pave the way for significantly accelerating their calculation by interrelating different approaches.
\end{abstract}

\setcounter{page}{1}
\maketitle

\section{Introduction and Motivation}
\label{sec:Motivation}
In the context of high-energy physics (HEP) and quantum field theories (QFT) in general, Feynman diagrams play a crucial role to compute precise predictions. These graphs encode interactions among the building blocks of the different models, allowing a pictorial representation of complicated scattering or time-evolution processes. Furthermore, the presence of closed loops inside Feynman diagrams gives rise to the so-called Feynman integrals, which possess many fascinating mathematical properties. Still, the explicit calculation of Feynman integrals, especially at higher-loop orders, is challenging.

Over the past few years, several methods have been developed to tackle Feynman integral calculations \footnote{Since it is not the purpose of this article to provide a full review of all the powerful methods recently developed to tackle Feynman integral calculations, we refer the interested reader to Ref. \cite{Heinrich:2020ybq} and references therein.}. Novel developments based on intersection theory \cite{Cacciatori:2022mbi,Fontana:2023amt,Crisanti:2024vnd} and geometry have shown promising connections that could eventually lead to a more efficient calculation of Feynman integrals and diagrams.

In this direction, the Loop-Tree Duality (LTD) \cite{Catani:2008xa,Rodrigo:2008fp,Bierenbaum:2010cy,Bierenbaum:2012th,Buchta:2014dfa,Tomboulis:2017rvd,Buchta:2015wna} is a formalism that relates loops with tree-level like topologies, thus writing in a different way the Feynman integrals. This idea was crucial to achieve a fully local (i.e. integrand-level) cancellation of infrared singularities in the so-called Four Dimensional Unsubtraction (FDU) formalism \cite{Hernandez-Pinto:2015ysa,Sborlini:2016gbr,Sborlini:2016hat}. 

Furthermore, it turns out that the application of LTD to multiloop diagrams has very interesting properties. One of the most important ones is that LTD representation is manifestly causal, in spite of requiring some re-writing or handling of the formulae in order to achieve a more simplified and explicit result. To be specific, when adding together all the terms of the representation, very compact results are obtained. These compact expressions lead to the so-called causal LTD representation \cite{Verdugo:2020kzh,Ramirez-Uribe:2020hes,Aguilera-Verdugo:2020kzc,Aguilera-Verdugo:2020nrp,Runkel:2019yrs,Runkel:2019zbm,Capatti:2019edf,Capatti:2019ypt,Capatti:2022mly} (or simply causal representation), which only contains terms with non-spurious or physical singularities. Each term of the causal representation is known as causal entangled threshold, since these are the building blocks of the singular structures present in any multiloop Feynman diagram. Recently, it was found that causal representation offers a powerful framework to calculate physical observables and cross-sections at higher-orders, starting from vacuum diagrams \cite{Ramirez-Uribe:2024rjg,LTD:2024yrb,Rodrigo:2024vda,deLejarza:2024scm}.

Besides brute force calculations, the causal representation can be obtained by using algebraic \cite{TorresBobadilla:2021dkq,TorresBobadilla:2021ivx} or {graph-based} \cite{Sborlini:2021owe} methods. Interestingly, it can be shown that the causal representation is \emph{not unique} in general. In this article, we exploit the graph-theoretic reconstruction approach to explore this degeneration. In order to do that, we analyze the symmetries among the different causal entangled thresholds and the graphs representing them. Then, we deepen into the kinematic constraints to generate equations relating different causal entangled thresholds. We combine all this information in order to understand the origin of the degeneration in the explicit formula for causal representations. {To facilitate the presentation, we use the one-loop case as a guiding line, as it provides a sufficiently illustrative framework for establishing the main discussion of this article.}

The outline of this paper is the following. In Sec. \ref{sec:CausalLTD} we explain with more details the LTD formalism, {and we depict the general graph-theoretic selection rules to reconstruct causal representations, emphazising the special constrains required for one-loop topologies. Additionally, in App. \ref{app:LTD}, we provide more details about the LTD-related terminology}. Then, in Sec. \ref{sec:BoxExample}, we use the box diagram to show explicitly that causal representation is degenerate, and we motivate the origin of this degeneration. After that, we move to a more complex example in Sec. \ref{sec:PentagonExample}, the pentagon diagram, and generalize the analysis to $N$-point one-loop Feynman diagrams in Sec. \ref{sec:NPointsOneLoop}. In Sec. \ref{sec:GeometryAlgebra}, we discuss the connection among the degeneration and the constraints relating different causal thresholds, {centering the discussion in the one-loop case. There, we focus on the interplay between symmetries of the graph and the number of algebraic constraints, both for minimally connected (mCG) and maximally connected graphs (MCG) in Sec. \ref{ssec:Comparison}}. Finally, we present the conclusions and future research directions in Sec. \ref{sec:Conclusions}.

\section{Causal Loop-Tree Duality}
\label{sec:CausalLTD}
As stated in the Introduction, the Loop-Tree Duality (LTD) transforms loop diagrams into collections of tree-level-like objects. This is achieved by applying Cauchy’s residue theorem to eliminate one degree of freedom per loop, specifically, by integrating the energy component of each loop momentum, which converts the integration domain from Minkowski to Euclidean space. Iterative application of the residue theorem directly leads to the dual representation. 

Furthermore, it turns out that adding together all the dual contributions, several simplifications occur directly at integrand level. The result, called the causal representation \cite{TorresBobadilla:2021dkq,TorresBobadilla:2021ivx,Sborlini:2021owe,Capatti:2022mly} is explicitly free of non-physical non-causal singularities. In detail, let us consider a generic $L$-loop scattering amplitude with $N$ external particles. The iterated application of nested residues and the subsequent combination of all dual contributions lead to
\beqn 
\nonumber {\cal A}_N^{(L)} &=& \sum_{\sigma \in \Sigma} \int_{\vec{\ell}_1 \ldots \vec{\ell}_L} \frac{{\cal N}_{\sigma}(\{q_{r,0}^{(+)}\},\{p_{j,0}\})}{x_{L+k}} \, \prod_{i=1}^k \frac{-1}{\lambda_{\sigma(i)}}
\\ &+& (\sigma \longleftrightarrow \bar{\sigma}) \, ,
\label{eq:causalRepresentation}
\eeqn
which is \emph{a} causal representation. In this expression, ${\cal N}$ is the numerator (which depends on the loop and external momenta), the integration measure is codified in
\beq
x_{L+k} = \prod_{j=1}^{L+k} 2 q^{(+)}_{j,0} \, ,
\label{eq:Prefactor}
\eeq
which is analogous to a phase-space integration measure and $1/\lambda_i$ are the causal propagators, where each $\lambda_i$ is associated with a physical threshold singularity of the diagram. {$k$ is the topological order of the reduced Feynman diagram \footnote{A reduced Feynman diagram is a graph in which all the propagators connecting the same vertices are collapsed into the so-called \emph{multi-edges}. For example, the well-known two-loop sunrise diagram is equivalent to a Feynman graph with a single multi-edge, i.e. \emph{without topological loops}. For a deeper discussion about this we refer the reader to Refs. \cite{Sborlini:2021owe,Ramirez-Uribe:2021ubp}.}, and is given by $k=V-1$, with $V$ the number of interaction vertices. Besides, $p_{i,0}$ is the energy of the external particle $i$ and $q_{j,0}^{(+)}$ is the positive on-shell energy associated to the multi-edge $j$ in the reduced Feynman diagram, which is defined as the positive solution of
\beq
q_{j}^2-m_j^2+\imath 0 \equiv 0 \, .
\eeq}
The symbol $\sigma$ represents a combination of $k$ \emph{entangled} causal thresholds from the set $\Sigma$ of all compatible causal thresholds. {In App. \ref{app:LTD} we deepen into these definitions for those readers that are not familiar with LTD notation.} 

For the sake of simplicity, we remove the prefactor $1/x_{L+k}$ since it is globally multiplying all the terms in the causal representation. Then, Eq. (\ref{eq:causalRepresentation}) can be re-written as
\beqn
\nonumber {\cal A}_N^{(L)} &=& \int_{\vec{\ell}_1 \ldots \vec{\ell}_L} \, \frac{A_{(N,{\rm RED})}^{(L)}(\{q_{i,0}^{(+)}\}_{i=1,\ldots,L+k},\{p_j\}_{j=1,\ldots,N})}{x_{L+k}} 
\\ &+& (\sigma \longleftrightarrow \bar{\sigma})\, ,
\label{eq:causalRepresentationReduced}
\eeqn
where we introduced the \emph{reduced amplitude} $A^{(L)}_{(N,{\rm RED})}$. We recall that this redefinition was useful to simplify the local renormalization algorithm introduced in Ref. \cite{Rios-Sanchez:2024xtv}.


\subsection{Review of graphic reconstruction}
\label{ssec:GeometryReview}
As we anticipated in the Introduction, the causal representation is not unique. Thus, we explore its degeneration and origin using the graph-theoretic properties of the underlying Feynman diagrams. 

In order to carry out this analysis, we make use of graphic selection rules to obtain the set of allowed causal-entangled thresholds, as described in Ref. \cite{Sborlini:2021owe}. Recall that causal thresholds $\lambda_i$ are in a one-to-one correspondence with connected binary partitions of the reduced Feynman graph with default outgoing external particles. For this reason, we can label causal thresholds by one of the sets of vertices determining the partition: by convention, we choose the smallest one taking into account the lexicographic ordering of vertices. As an example of the last sentence, if we have a box-graph $G=(V,E)$ with $V=\{v_1,v_2,v_3,v_4\}$, we denote $\lambda\equiv\{v_1,v_2\}$ instead of $\lambda\equiv\{v_3,v_4\}$ even if both are related to the same partition \footnote{Using group theory language, lexicographic ordering is used as a condition to select the representative element of the set of vertices associated to a given causal threshold o binary connected partition.}.

Also, the orientation or signature of a given causal threshold is related to the direction of the external legs w.r.t. the internal momenta flow. The convention that we follow is
\begin{equation}
    \lambda^{\pm} = \sum_{i \in I} q_{i,0}^{(+)} \pm \sum_{j \in O} p_{j,0} \, ,
    \label{eq:LambdaDefinition}
\end{equation}
{where $I$ is the set of internal propagators that are being cut and $O$ is the set of external momenta included in one of the elements of binary partition associated to $\lambda$. From a graphic point of view, $\lambda^+$ ($\lambda^-$) is used when the external and the cut internal momenta are aligned in the same (opposite) direction.} Notice that this definition is unambiguous since external particles must fulfill momentum conservation.

\begin{figure}[ht]
\begin{center}
\includegraphics[scale=0.11]{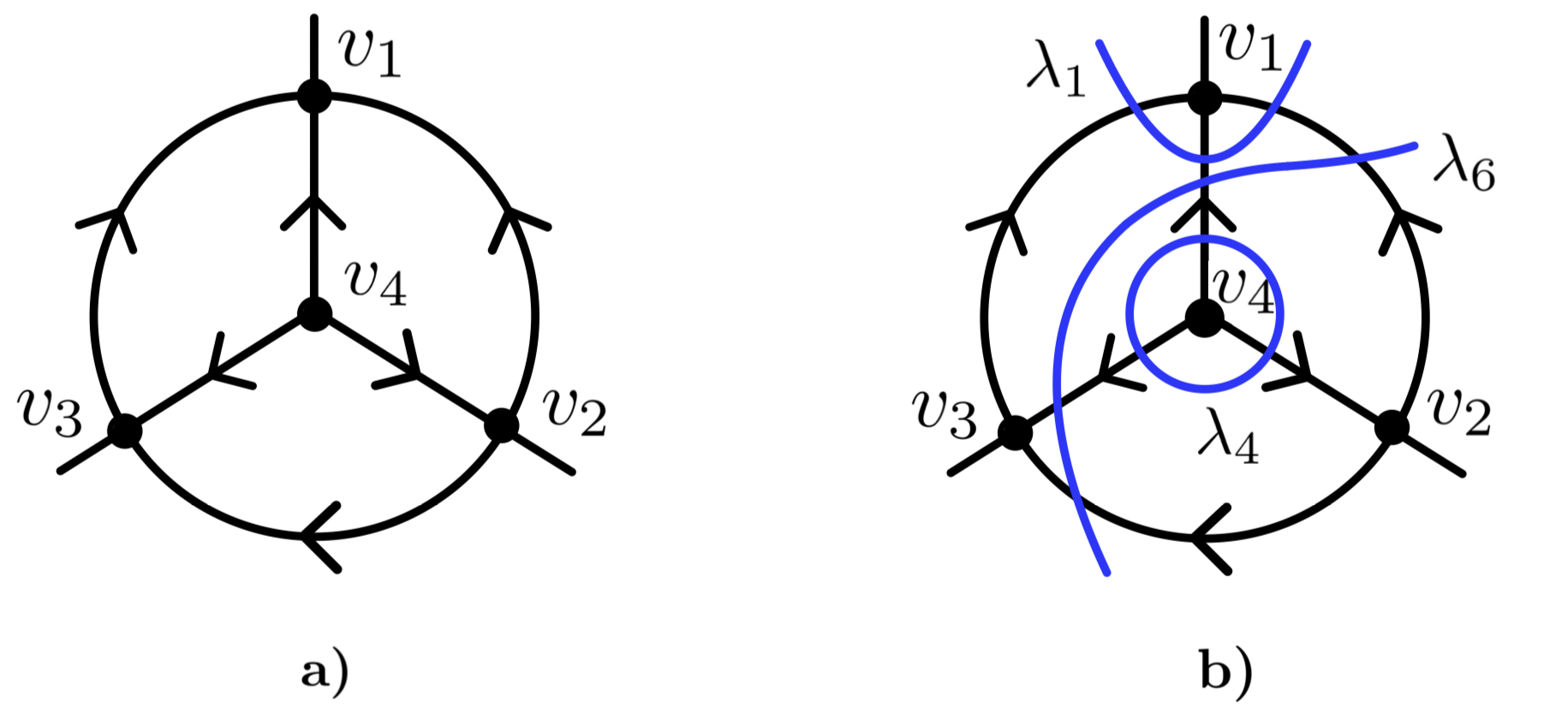}
\caption{Causal bootstrapping of terms contribution to the causal representation of Mercedes-Benz diagram. (a) First, we order the internal lines to avoid configurations with loops. (b) Then, we draw causal thresholds compatible with the fixed orientations of the internal lines.}
\label{fig:Vestido}
\end{center}
\end{figure}

After these introductory concepts, the set $\Sigma$ can be derived from a purely {graph-theoretic} formulation, {which is similar to the well-known Cutkosky rules \cite{Cutkosky:1960sp} and the ones described in \texttt{Diagrammar} \cite{tHooft:1973wag,tHooft:1974kac}}. The selection rules for allowed combinations of causal thresholds are:
\begin{enumerate}
\setcounter{enumi}{0}
    \item \emph{Consistent multi-edge momenta orientation}: {The orientations of the energy flow through the edges that cross a given cut should all be consistent, namely flow in the same direction.}
    \item \emph{All the multi-edges are cut by, at least, one of the causal thresholds inside the entangled combination}.
    \item \emph{No crossing of thresholds}: Two compatible causal propagators $\lambda_p$ and $\lambda_q$ must fulfill that the associated sets of connected vertices are disjoint or one is totally included in the other. Graphically, this means that the lines representing the cuts do no intersect each other.
\end{enumerate}
The first rule is equivalent to identifying all the possible directed acyclic graphs (DAG) of the corresponding reduced Feynman diagram, as we showed in Ref. \cite{Ramirez-Uribe:2021ubp,Clemente:2022nll}. So, we can first identify the DAGs and then dress them with suitable combinations of $\lambda$'s that meet conditions 2 and 3. This procedure is depicted in Fig. \ref{fig:Vestido}, with an example of a three-loop Mercerdes-Benz diagram. After we orient the internal multi-edges to avoid having loops (i.e. we deal with a DAG), we need to draw three $\lambda$'s on top of the graph.

{Since causal thresholds $\lambda$ are related to cuts of the diagram, selection rule 3 implies a compatibility constraint when computing sequential \emph{cut integrals}, as discussed in Ref. \cite{Abreu:2014cla}. In that article, it is shown that cuts of distinct channels must be compatible with iterated discontinuities on the same channels. By assigning a distinct color to the vertices on each side of the partition, one enforces that every subsequent cut is taken between vertices sharing the same color. Although the rules for constructing the cut integrals are, in principle, different from those employed in causal LTD for reconstructing the complete integrand, they ultimately lead to equivalent results. A comprehensive analysis of this interesting finding is deferred for a future publication.}

{Based on previous studies \cite{Sborlini:2021owe,TorresBobadilla:2021dkq,TorresBobadilla:2021ivx}, we have solid evidence that these rules are enough to determine $\Sigma$ and bootstrap the causal representation for maximally connected graphs (MCG). However, as also reported in Ref. \cite{Sborlini:2021owe}, it turns out that selection rules 1-3 are not enough to directly bootstrap the causal representation for non-MCG}. With the phrase \emph{directly bootstrap} we mean that the elements of the set $\Sigma$ are such that Eq. (\ref{eq:causalRepresentation}) with ${\cal N}_\sigma=1$ is the causal representation of the corresponding scalar Feynman graph. Instead, we find that an additional selection rule is required. It codifies the fact that external legs contain also geometrical information that constrains the simultaneously allowed causal entangled thresholds. For $N$-point one-loop amplitudes, we have to impose:
\begin{enumerate}
\setcounter{enumi}{3}
    \item \emph{Consistent external leg orientation}: We assume all the external legs to be outgoing, except for one which enters the diagram. Then, we apply the consistent alignment criterion (selection rule 1) and retain only those configurations fulfilling it.
\end{enumerate}
{We want to highlight that this rule was heuristically found, and its meaning will be carefully discussed in this work since it is connected with the origin of the degeneration of causal representations. {It is motivated by the fact that MCG have a well-defined causal representation and \emph{a} causal representation for any arbitrary graph can be obtained by sequentially removing internal lines. By imposing constraints on the external legs, rule 4 aims to remove the degeneration by joining vertices that were originally connected in the parent MCG. This procedure mirrors the method outlined in Ref. \cite{TorresBobadilla:2021dkq}.}

Also,} notice that choosing one leg to be \emph{incoming} immediately breaks the rotational invariance of $N$-points one-loop graphs: this observation is a key aspect of our analysis.


\section{Analysis of the causal representation of a box}
\label{sec:BoxExample}
\begin{figure}[ht]
\begin{center}
\includegraphics[scale=0.12]{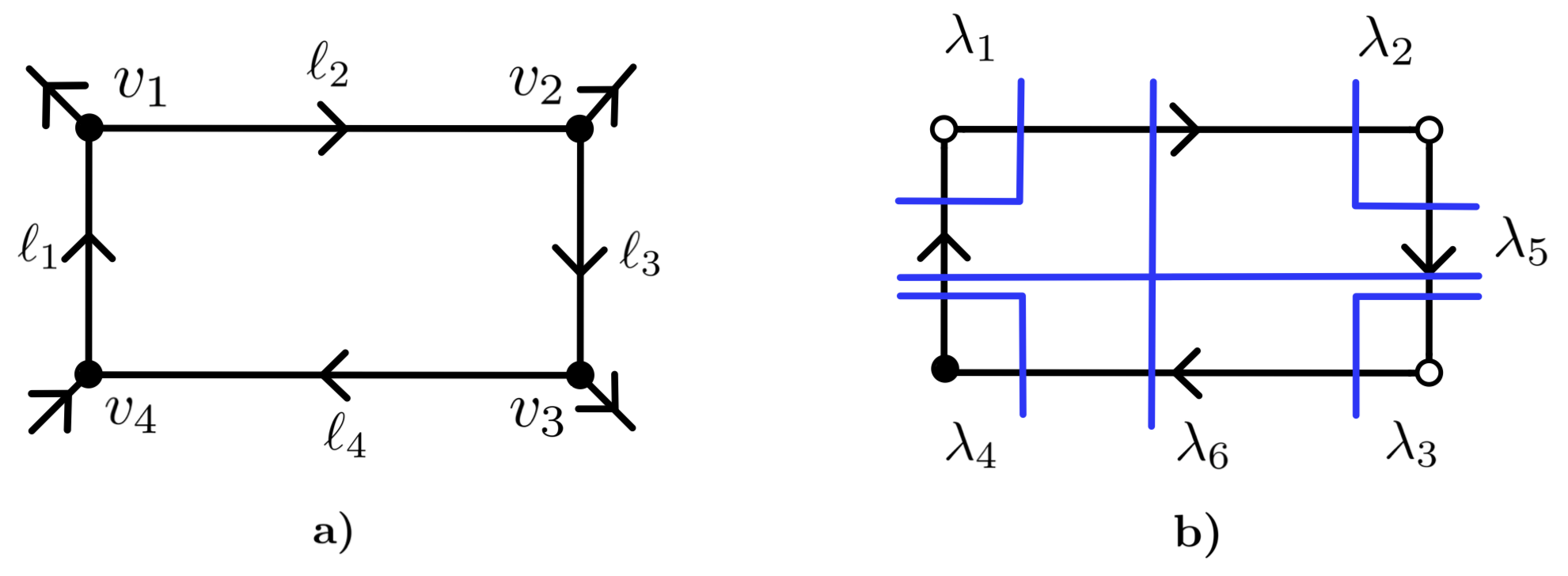}
\caption{Reference configuration for studying the one-loop box diagram. (a) Edge and vertex labeling, including the convention for outgoing external momenta. We choose $p_4$ to be incoming in order to satisfy momentum conservation for positive-energy particles. (b) graphic description of the possible causal thresholds $\lambda_i$.}
\label{fig:Box1}
\end{center}
\end{figure}

First, let us study the simplest non-trivial case: the one-loop box. In Fig. \ref{fig:Box1} we depict the diagram, labeling the edges {(with arbitrary orientations for reference purposes only)} and vertices, as well as we indicate the causal thresholds $\lambda_i$. 

The causal thresholds are
\beqn 
\lambda_i^{\pm} &=& q_{i,0}^{(+)} + q_{i+1,0}^{(+)} \pm p_{i,0} \equiv \{v_i\} \ {\rm for} \  i\in[1\ldots 4]\, ,
\label{eq:LambdasSimplesBOX}
\\ \lambda_5^{\pm} &=& q_{1,0}^{(+)}+q_{3,0}^{(+)} \pm (p_{1,0}+p_{2,0}) \equiv \{v_1,v_2\} \,  ,
\\ \lambda_6^{\pm} &=& q_{2,0}^{(+)}+q_{4,0}^{(+)} \pm (p_{2,0}+p_{3,0}) \equiv \{v_1,v_4\} \,  ,
\eeqn
with $q_{i,0}^{(+)}$ the positive on-shell energy corresponding to edge $e_i$, $i+1=5 \equiv 1$ and $p_4=-p_1-p_2-p_3$ by momentum conservation. Relying on the algorithms described in Refs. \cite{Sborlini:2021owe,IRENETFG}, \emph{a} causal representation can be written as:
\beq
A_{(4,{\rm RED})}^{(1)} = {\cal P}_{4}^{\rm Inv.} + {\cal P}_{4}^{\rm Deg.}(1) \, ,
\label{eq:BoxCausal1}
\eeq
where
\beqn
\nonumber {\cal P}_{4}^{\rm Inv.} &=&
\frac{1}{\lambda^{-}_3 \lambda^{-}_6 \lambda^{+}_1}+\frac{1}{\lambda^{-}_3 \lambda^{+}_1 \lambda^{+}_5}+\frac{1}{\lambda^{-}_3 \lambda^{+}_2 \lambda^{+}_5}
\\ \nonumber &+& \frac{1}{\lambda^{-}_2 \lambda^{-}_6 \lambda^{+}_1} + \frac{1}{\lambda^{+}_1 \lambda^{-}_4 \lambda^{+}_5} + \frac{1}{\lambda^{+}_2 \lambda^{-}_4 \lambda^{+}_5}
\\ &+& \frac{1}{\lambda^{+}_2 \lambda^{-}_4 \lambda^{+}_6} + \frac{1}{\lambda^{+}_3 \lambda^{-}_4 \lambda^{+}_6} \, ,
\label{eq:BoxInvariante}
\eeqn
and
\beqn
{\cal P}_{4}^{\rm Deg.}(1) &=&
\frac{1}{\lambda^{-}_2 \lambda^{+}_1 \lambda^{+}_3}+\frac{1}{\lambda^{+}_1 \lambda^{+}_3 \lambda^{-}_4} \, ,
\label{eq:BoxDegenerado1}
\eeqn
corresponds to the \emph{invariant} and \emph{degenerate} parts, respectively, whose precise meaning will be discussed in the next subsection. The result given in Eq. (\ref{eq:BoxCausal1}) agrees with previously reported causal representations for the one-loop box \cite{TorresBobadilla:2021dkq,TorresBobadilla:2021ivx,Sborlini:2021owe}.


\subsection{Degeneration and rotational symmetry}
\label{ssec:rotational}
The straightforward application of selection rules 1-3 to the box diagram leads to the following set of allowed causal entangled thresholds:
\beqn
\nonumber \Sigma_{4} &=& \{\{1,2,3\},\{1,2,4\},\{1,2,6\},\{1,3,4\},\{1,3,5\},
\\ \nonumber && \{1,3,6\},\{1,4,5\},\{2,3,4\},\{2,3,5\},\{2,4,5\},
\\ && \{2,4,6\},\{3,4,6\}\} \, ,
\eeqn
where we are using the short-hand notation 
\beq
\{\sigma(1),\ldots,\sigma(k)\} \equiv \lambda_{\sigma(1)} \cdot \ldots \cdot \lambda_{\sigma(k)} \, ,
\eeq
and $k$ denotes the topological order of the diagram. We omit the signature of each causal threshold since it is unambiguously defined by selection rule 1 and the convention of external momenta directions depicted in Fig. \ref{fig:Box1}. 

Notice that $\Sigma_4$ contains 12 elements. However, if we try to reconstruct the amplitude using Eq. (\ref{eq:causalRepresentation}) with $\Sigma_4$, it fails. This is because we need to implement the 4-th rule, which eliminates two configurations, i.e.
\beq
\{\{1,2,4\},\{2,3,4\}\} \, ,
\eeq
and leads to
\beqn
\nonumber \bar{\Sigma}_{4} &=& \{\{1, 2, 3\}, \{1, 2, 6\}, \{1, 3, 4\}, \{1, 3, 5\},  
\\ \nonumber && \{1, 3, 6\}, \{1, 4, 5\}, \{2,
   3, 5\}, \{2, 4, 5\}, 
\\ && \{2, 4, 6\}, \{3, 4, 6\}\} \, ,
\eeqn
which allows to directly reconstruct the integrand-level representation of the box according to Eq. (\ref{eq:causalRepresentation}).

\begin{figure}[ht]
\begin{center}
\includegraphics[scale=0.128]{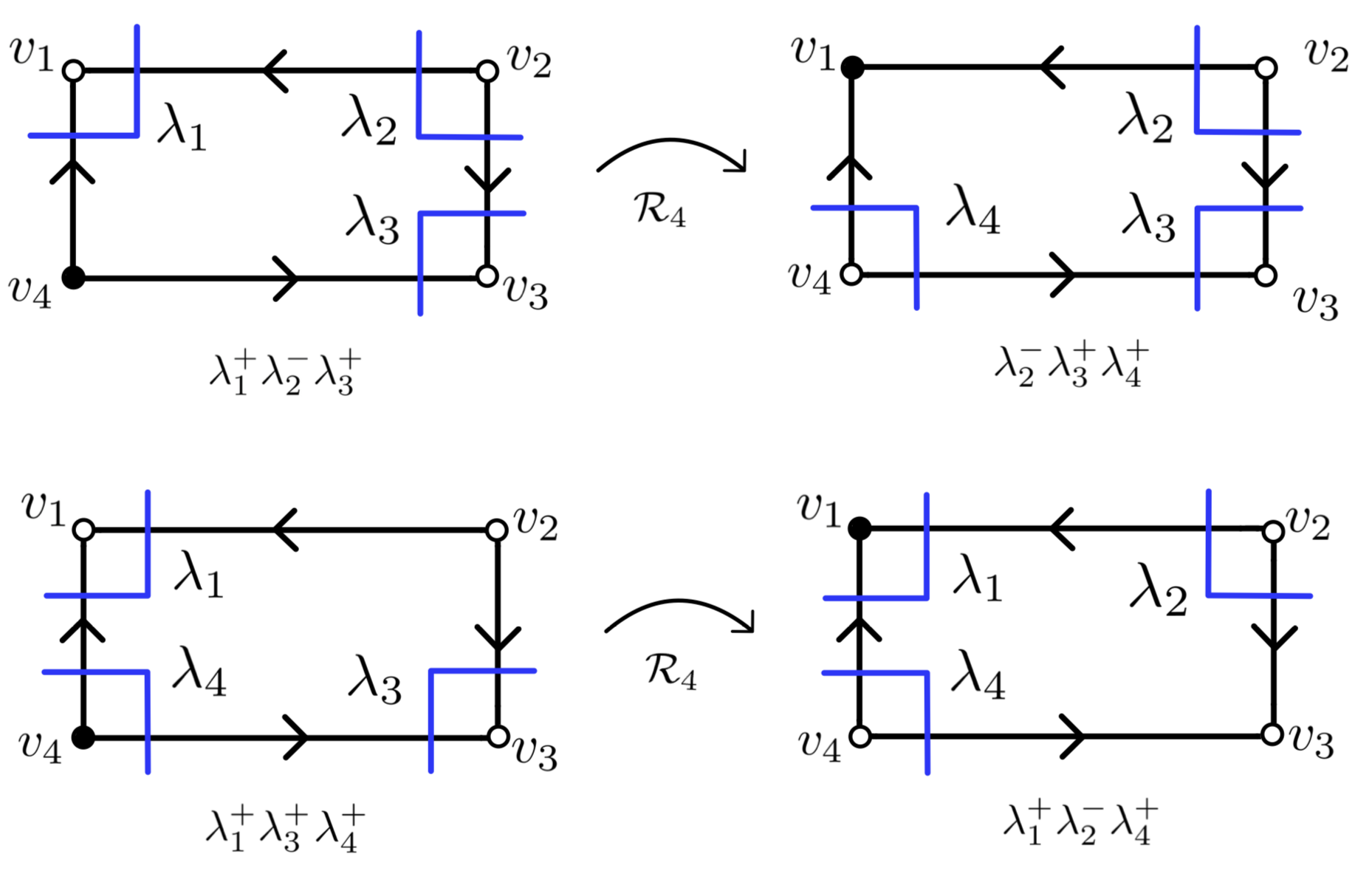}
\caption{Application of the ${\cal R}_4$ transformation to elements in $\bar{\Sigma}_4$ not present in $\Sigma_4$.}
\label{fig:BoxRotation}
\end{center}
\end{figure}

At this point, let us stop and think about the meaning of the fourth selection rule. We mentioned that, by convention, all the external particles are considered \emph{outgoing}. The system is constrained by momentum conservation, thus,
\beq
\sum_{i=1}^4 \, p_{i,0} = 0 \, ,
\eeq
and particles are required to be real, which implies $p_{i,0} \geq 0$. Then at least one of the external particles must be chosen incoming: in our example, $p_4=-p_1-p_2-p_3$. When we do this, one of the vertices is automatically selected as \emph{the chosen one} and we can define some transformations of the diagram. {For instance, we can rotate the vertices according to:
\beq
{\cal R}_N = \{v_i \to v_{i+1} \quad ({\rm mod}\, N)\ \} \, ,
\eeq
with $N$ the number of legs (or vertices in the one-loop case only). This transformation can also be understood as acting on the causal thresholds, since they are defined as subsets of vertices. As an example, if we consider $\lambda_5=\{v_1,v_2\}$ then
\beq
{\cal R}_4(\lambda_5) = \{v_2,v_3\} \equiv \{v_1,v_4\} = \lambda_6 \, , 
\eeq
where we relied on the fact that $\{v_2,v_3\}$ and $\{v_2,v_3\}$ represent the same binary connected partition of $\{v_1,v_2,v_3,v_4\}$. Abusing of the notation, ${\cal R}_4$ also acts on causal entangled thresholds, $\sigma$, transforming each individual causal threshold $\lambda_{\sigma(i)}$ with $i\in[1\ldots k]$.

Returning to the analysis of the box}, if we draw the removed configurations and apply ${\cal R}_4$ we obtain:
\beq
{\cal R}_4 \left(\{\{1,2,3\},\{1,3,4\}\}\right) = \{\{1,2,4\},\{2,3,4\}\} \, ,
\eeq
as we depict in Fig. \ref{fig:BoxRotation}. {It is important to note that ${\cal R}_4$ acts on the causal representation (that is, on the structure of the various $\lambda$’s) without altering the momentum labels. The black vertex in Fig.~\ref{fig:BoxRotation} marks the \emph{chosen vertex} and serves as the reference for rotating the causal thresholds that dress the diagram (i.e. the blue lines associated with the cuts). Because the vertex labels remain unchanged, $p_4$ is always incoming, whereas the remaining external momenta are outgoing. Consequently, in the first row on the left, $\lambda_3^{+}$ is transformed into $\lambda_4^{+}$, as the internal and external momenta are aligned in the same direction. The signature obtained after applying ${\cal R}_4$ is therefore fully consistent with the original labeling and orientation of the external momenta. 

Thus, }in terms of contributions to the causal representation, we have
\beqn
\nonumber {\cal R}_4\left({\cal P}_{4}^{\rm Deg.}(1)\right) &=&
\frac{1}{\lambda^{-}_3 \lambda^{+}_4 \lambda^{+}_2}+\frac{1}{\lambda^{-}_2 \lambda^{+}_1 \lambda^{-}_4} 
\\ &=& {\cal P}_{4}^{\rm Deg.}(2)\, .
\label{eq:BoxDegenerado2}
\eeqn
This justifies the use of the word \emph{degenerate}: both ${\cal P}_{4}^{\rm Deg.}(1)$ and ${\cal P}_{4}^{\rm Deg.}(2)$ can be used to write the causal representation, since they are totally equivalent.

Regarding the other contribution, ${\cal P}_4^{\rm Inv.}$, if we apply the transformation ${\cal R}_4$ to its elements, we recover the same set: the subset of causal entangled thresholds is invariant under the transformation.


\subsection{Symmetric causal representation}
\label{ssec:Symetrization}
Let us examine more thoroughly the mechanisms underlying the transformation ${\cal R}_4$. First, let us write
\beq
\Sigma_4 = \Sigma_4^{\rm Inv.} \cup \Sigma_4^{\rm Deg.}(1) \cup \Sigma_4^{\rm Deg.}(2) \, ,
\label{eq:degenerateS4}
\eeq
and notice that
\beqn
{\cal R}_4\left(\Sigma_4^{\rm Inv.}\right) &\subseteq& \Sigma_4^{\rm Inv.} \, ,
\\ {\cal R}_4\left(\Sigma_4^{\rm Deg.}(i)\right) &\subseteq& \Sigma_4^{\rm Deg.}(j) \quad i\neq j \, ,
\eeqn
with $i,j \in \{1,2\}$. Also, any rotation operator fulfills
\beq
({\cal R}_N)^N = I \, ,
\eeq
which implies that if we apply it $N$ times to any element, we get the same thing again. 

Another observation somewhat obvious but very important is that applying the rotation operator to a causal representation leads to another causal representation. Having said that, 
\beq
{\cal R}_4^{i} \left({\cal P}_4^{\rm Inv.} + {\cal P}_4^{\rm Deg.}(1)\right) 
\eeq
is a causal representation for any $i \in \{1,2,3,4\}$.

Then, we propose a \emph{first algorithm} to generate a \emph{symmetrized causal representation}. We start from $\bar{\Sigma}$ and we calculate the set of elements connected by successive rotations, i.e.
\beqn
O\left[{\cal R}_N\right](\sigma) &=& \{\sigma,{\cal R}_N(\sigma),\ldots,{\cal R}^{N-1}_N(\sigma)\} \, ,
\eeqn
for each $\sigma \in \bar{\Sigma}$. Abusing a bit of the notation, identifying the contribution to the integrand-level causal representation with the causal entangled thresholds, i.e.
\beq 
\sigma \equiv \prod_{i=1}^k \lambda_{\sigma(i)}^{-1} \, ,
\eeq
we define
\beq
A_{(N,{\rm RED})}^{(1),{\rm Sym.}} = \frac{1}{N} \, \sum_{\sigma \in \bar{\Sigma}} O\left[{\cal R}_N\right](\sigma) \, ,
\label{eq:Simetrizada1}
\eeq
that is also a causal representation. We notice that if $\sigma \in \Sigma_N^{\rm Inv.}$, then the element appears $N$ times in the sum present in Eq. (\ref{eq:Simetrizada1}), but if $\sigma \in \Sigma_N^{\rm Deg.}(j)$ its multiplicity is smaller.

Applying this algorithm to the particular case of the box diagram, we find
\beqn
\nonumber A_{(4,{\rm RED})}^{(1),{\rm Sym.}} &=& \frac{1}{2} \left( \frac{1}{\lambda^{-}_3 \lambda^{+}_4 \lambda^{+}_2}+\frac{1}{\lambda^{-}_2 \lambda^{+}_1 \lambda^{-}_4} \right.
\\ \nonumber &+& \left.  \frac{1}{\lambda^{-}_2 \lambda^{+}_1 \lambda^{+}_3} + \frac{1}{\lambda^{+}_1 \lambda^{+}_3 \lambda^{-}_4} \right)
\\ \nonumber &+& \frac{1}{\lambda^{-}_3 \lambda^{-}_6 \lambda^{+}_1}+\frac{1}{\lambda^{-}_3 \lambda^{+}_1 \lambda^{+}_5}+\frac{1}{\lambda^{-}_3 \lambda^{+}_2 \lambda^{+}_5}
\\ \nonumber &+& \frac{1}{\lambda^{-}_2 \lambda^{-}_6 \lambda^{+}_1} + \frac{1}{\lambda^{+}_1 \lambda^{-}_4 \lambda^{+}_5} + \frac{1}{\lambda^{+}_2 \lambda^{-}_4 \lambda^{+}_5}
\\ &+& \frac{1}{\lambda^{+}_2 \lambda^{-}_4 \lambda^{+}_6} + \frac{1}{\lambda^{+}_3 \lambda^{-}_4 \lambda^{+}_6} \, .
\label{eq:SimetrizadaBox}
\eeqn
Notice also, that \emph{all} the elements of $\Sigma_4$ (i.e. the set resulting from applying only selection rules 1-3) are present in Eq. (\ref{eq:SimetrizadaBox}): those elements in ${\cal P}_4^{\rm Deg.}(i)$ have a symmetry factor $1/2$, whilst the ones in the invariant set are multiplied directly by 1.

Before moving forward to the next topology, let us notice that we can alternatively write $A_{(4,{\rm RED})}^{(1),{\rm Sym.}}$ as
\beq
A_{(4,{\rm RED})}^{(1),{\rm Sym.}} \equiv O\left[{\cal R}_N\right]\left(g(A_{4,{\rm RED}}^{(1)})\right) \, ,
\eeq
where we understand the sum of the elements that belong to the orbit of
\beqn
\nonumber g(A_{4,{\rm RED}}^{(1)}) &=&  \frac{1}{2} \left(\frac{1}{\lambda^{-}_2 \lambda^{+}_1 \lambda^{+}_3}+\frac{1}{\lambda^{+}_1 \lambda^{+}_3 \lambda^{-}_4} \right)
\\ \nonumber &+& \frac{1}{4} \left(\frac{1}{\lambda^{-}_3 \lambda^{-}_6 \lambda^{+}_1}+\frac{1}{\lambda^{-}_3 \lambda^{+}_1 \lambda^{+}_5}+\frac{1}{\lambda^{-}_3 \lambda^{+}_2 \lambda^{+}_5} \right.
\\ \nonumber &+& \left. \frac{1}{\lambda^{-}_2 \lambda^{-}_6 \lambda^{+}_1} + \frac{1}{\lambda^{+}_1 \lambda^{-}_4 \lambda^{+}_5} + \frac{1}{\lambda^{+}_2 \lambda^{-}_4 \lambda^{+}_5} \right.
\\ &+& \left. \frac{1}{\lambda^{+}_2 \lambda^{-}_4 \lambda^{+}_6} + \frac{1}{\lambda^{+}_3 \lambda^{-}_4 \lambda^{+}_6} \right)\, ,
\eeqn
considering this as a \emph{generator} of the symmetric causal representation. 


\section{Increasing the complexity: pentagon diagrams}
\label{sec:PentagonExample}
\begin{figure}[ht]
\begin{center}
\includegraphics[scale=0.12]{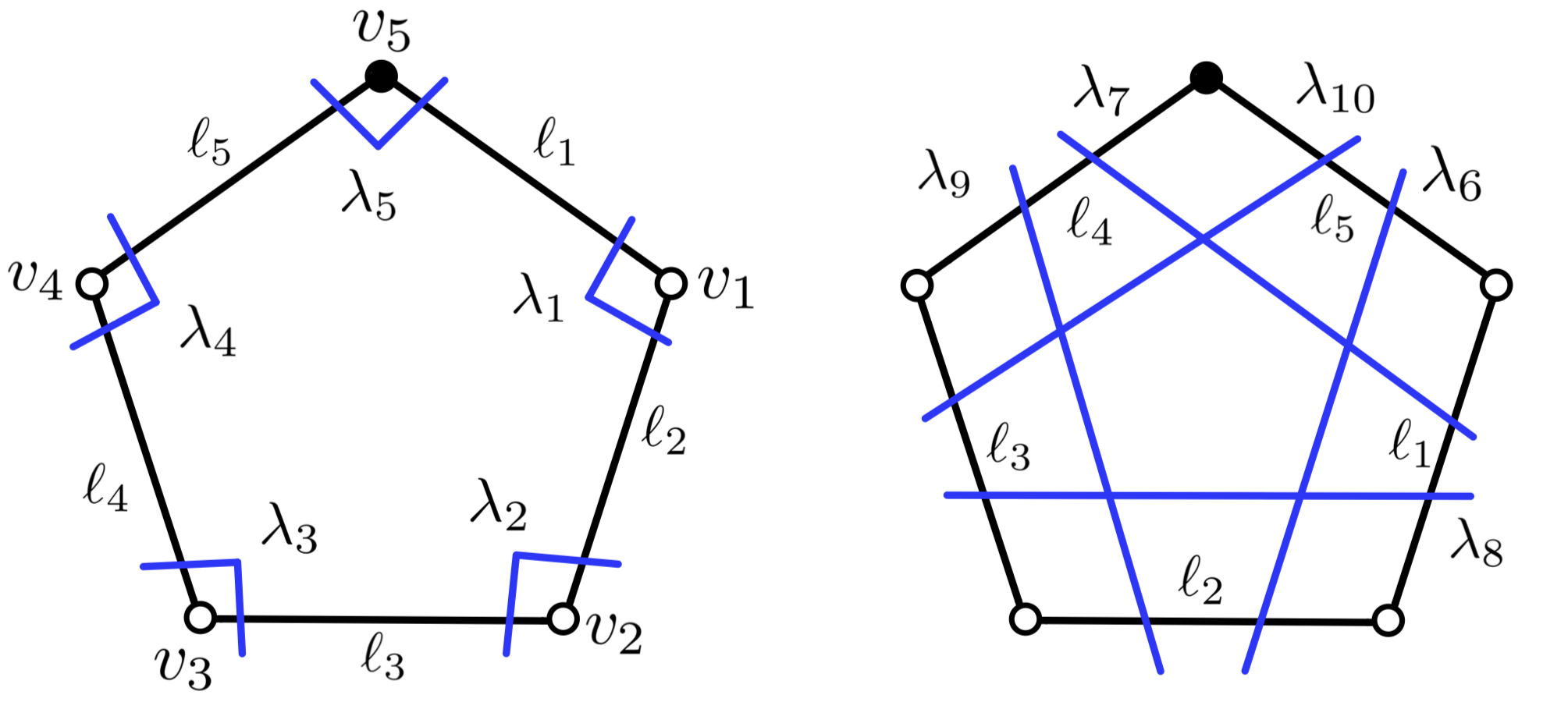}
\caption{Reference configuration for studying the one-loop pentagon diagram, including edge and vertex labeling, as well as a graphic description of the possible causal thresholds $\lambda_i$. We choose $p_5$ to be incoming in order to satisfy momentum conservation for positive-energy particles.}
\label{fig:Pentagon1}
\end{center}
\end{figure}

After studying in detail the box, let us move to a slightly more complex example: the pentagon. This time, as depicted in Fig. \ref{fig:Pentagon1}, there are 10 possible causal thresholds:
\beqn 
\lambda_i^{\pm} &=& q_{i,0}^{(+)} + q_{i+1,0}^{(+)} \pm p_{i,0} \equiv \{v_i\} \ \ {\rm for} \ i\in[1\ldots 5]\, ,
\label{eq:LambdasSimplesPENTAGON}
\\ \lambda_6^{\pm} &=& q_{1,0}^{(+)}+q_{3,0}^{(+)} \pm p_{12,0} \equiv \{v_1,v_2\} \,  ,
\\ \lambda_7^{\pm} &=& q_{2,0}^{(+)}+q_{5,0}^{(+)} \pm p_{234,0} \equiv \{v_1,v_5\} \,  ,
\\ \lambda_8^{\pm} &=& q_{2,0}^{(+)}+q_{4,0}^{(+)} \pm p_{23,0} \equiv \{v_2,v_3\} \,  ,
\\ \lambda_9^{\pm} &=& q_{3,0}^{(+)}+q_{5,0}^{(+)} \pm p_{34,0} \equiv \{v_3,v_4\} \,  ,
\\ \lambda_{10}^{\pm} &=& q_{1,0}^{(+)}+q_{4,0}^{(+)} \pm p_{123,0} \equiv \{v_4,v_5\} \,  ,
\eeqn
with $i+1=6\equiv 1$ and $p_{ijk,0}=p_{i,0}+p_{j,0}+p_{k,0}$. Applying selection rules 1-3, we find
\beqn 
\nonumber \Sigma_5 &=& \{\{1, 2, 3, 4\}, \{1, 2, 3, 5\}, \{1, 2, 3, 7\}, \{1, 2, 3, 9\}, 
\\ \nonumber && \{1, 2, 4, 5\}, \{1, 2, 4, 7\}, \{1, 2, 4, 8\}, \{1, 2, 4, 9\}, 
\\ \nonumber && \{1, 2, 4, 10\}, \{1, 2, 5, 8\}, \{1, 2, 5, 10\}, \{1, 2, 7, 8\}, 
\\ \nonumber && \{1, 3, 4, 5\}, \{1, 3, 4, 6\}, \{1, 3, 4, 7\}, \{1, 3, 4, 8\}, 
\\ \nonumber && \{1, 3, 4, 10\}, \{1, 3, 5, 6\}, \{1, 3, 5, 8\}, \{1, 3, 5, 9\}, 
\\ \nonumber && \{1, 3, 5, 10\}, \{1, 3, 6, 9\}, \{1, 3, 7, 8\}, \{1, 3, 7, 9\}, 
\\ \nonumber && \{1, 4, 5, 6\}, \{1, 4, 5, 9\}, \{1, 4, 6, 9\}, \{1, 4, 6, 10\}, 
\\ \nonumber && \{1, 4, 7, 9\}, \{1, 5, 6, 10\}, \{2, 3, 4, 5\}, \{2, 3, 4, 6\}, 
\\ \nonumber && \{2, 3, 4, 10\}, \{2, 3, 5, 6\}, \{2, 3, 5, 7\}, \{2, 3, 5, 9\}, 
\\ \nonumber && \{2, 3, 5, 10\}, \{2, 3, 6, 9\}, \{2, 4, 5, 6\}, \{2, 4, 5, 7\}, 
\\ \nonumber && \{2, 4, 5, 8\}, \{2, 4, 5, 9\}, \{2, 4, 6, 9\}, \{2, 4, 6, 10\}, 
\\ \nonumber && \{2, 4, 8, 10\}, \{2, 5, 6, 10\}, \{2, 5, 7, 8\}, \{2, 5, 8, 10\}, 
\\ \nonumber && \{3, 4, 5, 7\}, \{3, 4, 5, 8\}, \{3, 4, 8, 10\}, \{3, 5, 7, 8\}, 
\\ && \{3, 5, 7, 9\}, \{3, 5, 8, 10\}, \{4, 5, 7, 9\}\} \, ,
\label{eq:Sigma5total}
\eeqn
which contains 55 elements. Imposing the fourth selection rule, we get
\beqn
\bar{\Sigma}_5 &=& \{\{1, 2, 3, 4\}, \{1, 2, 3, 9\}, \{1, 2, 4, 7\}, 
\\ \nonumber && \{1, 2, 4, 8\}, \{1, 2, 4, 9\}, \{1, 2, 7, 8\}, \{1, 3, 4, 6\}, 
\\ \nonumber && \{1, 3, 4, 8\}, \{1, 3, 4, 10\}, \{1, 3, 5, 9\}, \{1, 3, 5, 10\}, 
\\ \nonumber && \{1, 3, 6, 9\}, \{1, 3, 7, 8\}, \{1, 3, 7, 9\}, \{1, 4, 5, 6\}, 
\\ \nonumber && \{1, 4, 5, 9\}, \{1, 4, 6, 9\}, \{1, 4, 6, 10\}, \{1, 4, 7, 9\}, 
\\ \nonumber && \{1, 5, 6, 10\}, \{2, 3, 4, 6\}, \{2, 3, 6, 9\}, \{2, 4, 5, 6\}, 
\\ \nonumber && \{2, 4, 5, 7\}, \{2, 4, 6, 9\}, \{2, 4, 6, 10\}, \{2, 4, 8, 10\}, 
\\ \nonumber && \{2, 5, 6, 10\}, \{2, 5, 7, 8\}, \{2, 5, 8, 10\}, \{3, 4, 8, 10\},
\\ \nonumber && \{3, 5, 7, 8\}, \{3, 5, 7, 9\}, \{3, 5, 8, 10\}, 
\\ && \{4, 5, 7, 9\}\} \, ,
\label{eq:Sigma5reduced}
\eeqn
where 20 elements were removed, and this set allows to reconstruct the pentagon amplitude according to Eq. (\ref{eq:causalRepresentation}).

Applying the symmetrization algorithm described in Eq. (\ref{eq:Simetrizada1}) we obtain
\beq
A_{(5,{\rm RED})}^{(1),{\rm Sym.}} =  \frac{1}{5} C^{(1)}_{5} + \frac{2}{5} C^{(2)}_{5} + \frac{3}{5} C^{(3)}_{5} + C^{(0)}_{5} \, ,
\label{eq:SimetrizadaPentagon}
\eeq
where the coefficients $C_{5}^{(i)}$ are reported in App. A. Again, all the elements of $\Sigma_5$ are present in this causal representation with the appropriated weight. In other words, they are all generated from the iterated application of ${\cal R}_5$ to elements of the standard causal representation. Also, notice that there are four different symmetry coefficients: $1/5$, $2/5$, $3/5$ and 1. This is an effect derived from the non-trivial intersection of $O[{\cal R}_N](\sigma)$ for each $\sigma \in \bar{\Sigma}_5$.

\begin{figure*}[ht]
\begin{center}
\includegraphics[scale=0.17]{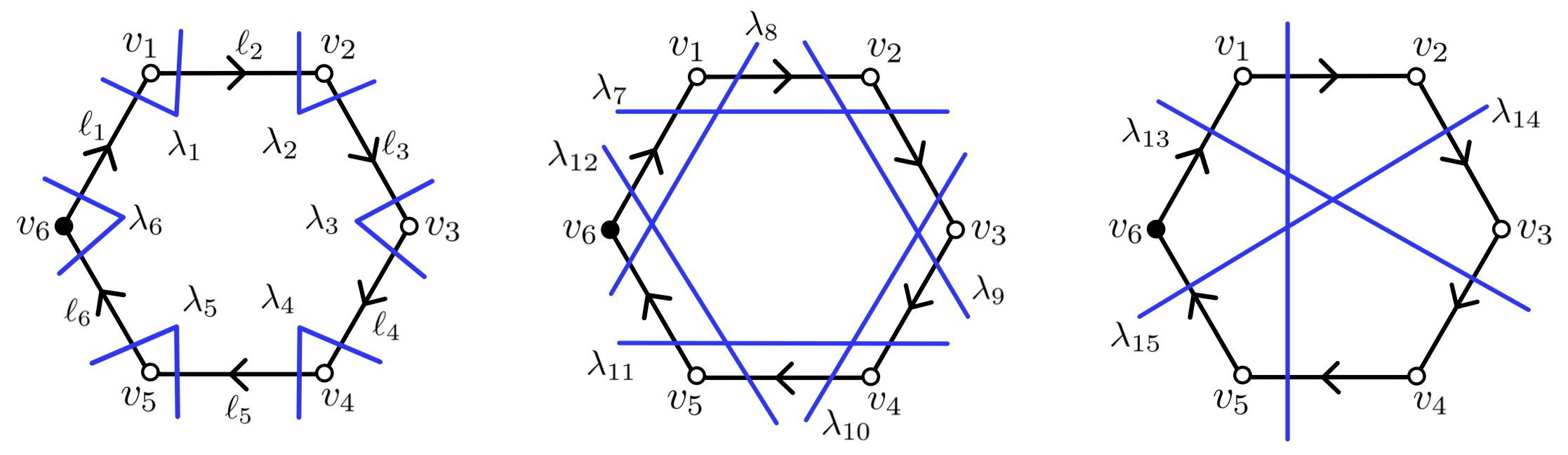}
\caption{Reference configuration for studying the one-loop hexagon diagram, including edge and vertex labeling, as well as a graphic description of the possible causal thresholds $\lambda_i$. We choose $p_6$ to be incoming in order to satisfy momentum conservation for positive-energy particles.}
\label{fig:Hexagon1}
\end{center}
\end{figure*}

\section{Towards general symmetries in one-loop $N$-point diagrams}
\label{sec:NPointsOneLoop}
After studying the box and the pentagon, it seems that we can generate a symmetric version of the causal representation by applying the rotation operator ${\cal R}_N$ to a given causal representation {for any $N$-point one-loop amplitude}. Moreover, this symmetric causal representation seems to be build directly from $\Sigma_N$, weighting each causal entangled threshold with an appropriated factor. So, two interconnected questions arise at this point:
\begin{enumerate}
    \item will the iterated application of ${\cal R}_N$ to a causal representation {for a one-loop amplitude} generate a symmetric one containing all the entangled thresholds in $\Sigma_N$ (i.e. those fulfilling selection rules 1-3)?
    \item and, is it possible to calculate the symmetry weight factors directly from $\Sigma_N$?
\end{enumerate}
Regarding the first question, we anticipate that the symmetrization of a causal representation does not always span the whole set $\Sigma_N$. Whilst providing an answer to the second question seems feasible, it is beyond the scope of the present work and is left for future research. 

Before moving forward in this section, we clarify that we will present only relevant selected parts of the results, due to the (very) cumbersome and lengthy formulae involved. {Also, we will only deal with the one-loop case: a more detailed analysis of multi-loop topologies is deferred for a future publication}. All the expressions, including the standard and symmetric causal representations as well as the dual representation from explicit residue calculation, are reported in an accompanying Zenodo repository \cite{ZENODO}.

Now, let us focus on the hexagon and apply the methodology deployed in the previous sections. We follow the conventions depicted in Fig. \ref{fig:Hexagon1}. This time, there are 15 causal thresholds:
\beqn
&& \lambda_i \equiv \{v_i\} \ \ {\rm for} \ i\in[1\ldots 6]\ ,
\label{eq:LambdasSimplesHEXAGON}
\\  && \lambda_7 \equiv \{v_1,v_2\} \,   ,  \ \lambda_8 \equiv \{v_1,v_6\} \,   ,  \ \lambda_9 \equiv \{v_2,v_3\} \,   ,
\\ &&  \lambda_{10} \equiv \{v_3,v_4\} \,   ,  \, \lambda_{11} \equiv \{v_4,v_5\} \,   ,  \, \lambda_{12} \equiv \{v_5,v_6\} \,   ,
\\  && \lambda_{13} \equiv \{v_1,v_2,v_3\} \,   ,  \ \lambda_{14} \equiv \{v_1,v_2,v_6\}  \,  ,
\\ && \lambda_{15} \equiv \{v_1,v_5,v_6\}  \, , 
\label{eq:LambdasHexagon}
\eeqn
where we use a short-hand notation to indicate only the vertex partitions involved. Applying selection rules 1-3 we obtain $\Sigma_6$ with 279 elements, and the application of the fourth rule leads to $\bar{\Sigma}_6$, with only 126 causal entangled thresholds. Then, the iterated application of ${\cal R}_6$ to the elements of $\bar{\Sigma}_6$ generates \emph{a symmetric} causal representation according to Eq. (\ref{eq:Simetrizada1}). However, the symmetry factors for some of the elements of $\Sigma_6$ are zero. This means that not all the causal entangled thresholds present in $\Sigma_6$ can be generated from rotations of $\bar{\Sigma}_6$, i.e.
\beq
O[{\cal R}_6](\bar{\Sigma}_6) \subset \Sigma_6 \, .
\eeq
This result is new for $N \geq 6$, and points towards additional transformations besides the simple rotation ${\cal R}_N$ and constrains imposed by selection rule 4.

\subsection{Analysis of symmetries in the hexagon}
\label{ssec:HexaAnalisis}
To understand what is going on, let us focus on the complement of $O[{\cal R}_6](\bar{\Sigma}_6)$ w.r.t. $\Sigma_6$. This set is defined by
\beqn
\nonumber && \left(O[{\cal R}_6](\bar{\Sigma}_6)\right)^{c} = \{ \{1, 2, 3, 4, 14\}, \{1, 2, 3, 5, 13\}, 
\\ \nonumber && \{1, 2, 3, 6, 15\}, \{1, 2, 4, 5, 13\}, \{1, 2, 4, 5, 14\}, 
\\ \nonumber && \{1, 2, 4, 6, 14\}, \{1, 2, 5, 6, 13\}, \{1, 3, 4, 5, 14\}, 
\\ \nonumber && \{1, 3, 4, 6, 14\}, \{1, 3, 4, 6, 15\}, \{1, 3, 5, 6, 15\}, 
\\ \nonumber && \{1, 4, 5, 6, 14\}, \{2, 3, 4, 5, 13\}, \{2, 3, 4, 6, 15\}, 
\\ \nonumber && \{2, 3, 5, 6, 13\}, \{2, 3, 5, 6, 15\}, \{2, 4, 5, 6, 13\}, 
\\ && \{3, 4, 5, 6, 15\} \} \, ,
\eeqn
and it cannot be connected to any element in $\bar{\Sigma}_6$. In other words, applying selection rule 4 by marking a \emph{distinguished vertex} associated to the incoming particle breaks a certain symmetry of the hexagon. 

\begin{figure*}[ht]
\begin{center}
\includegraphics[scale=0.2]{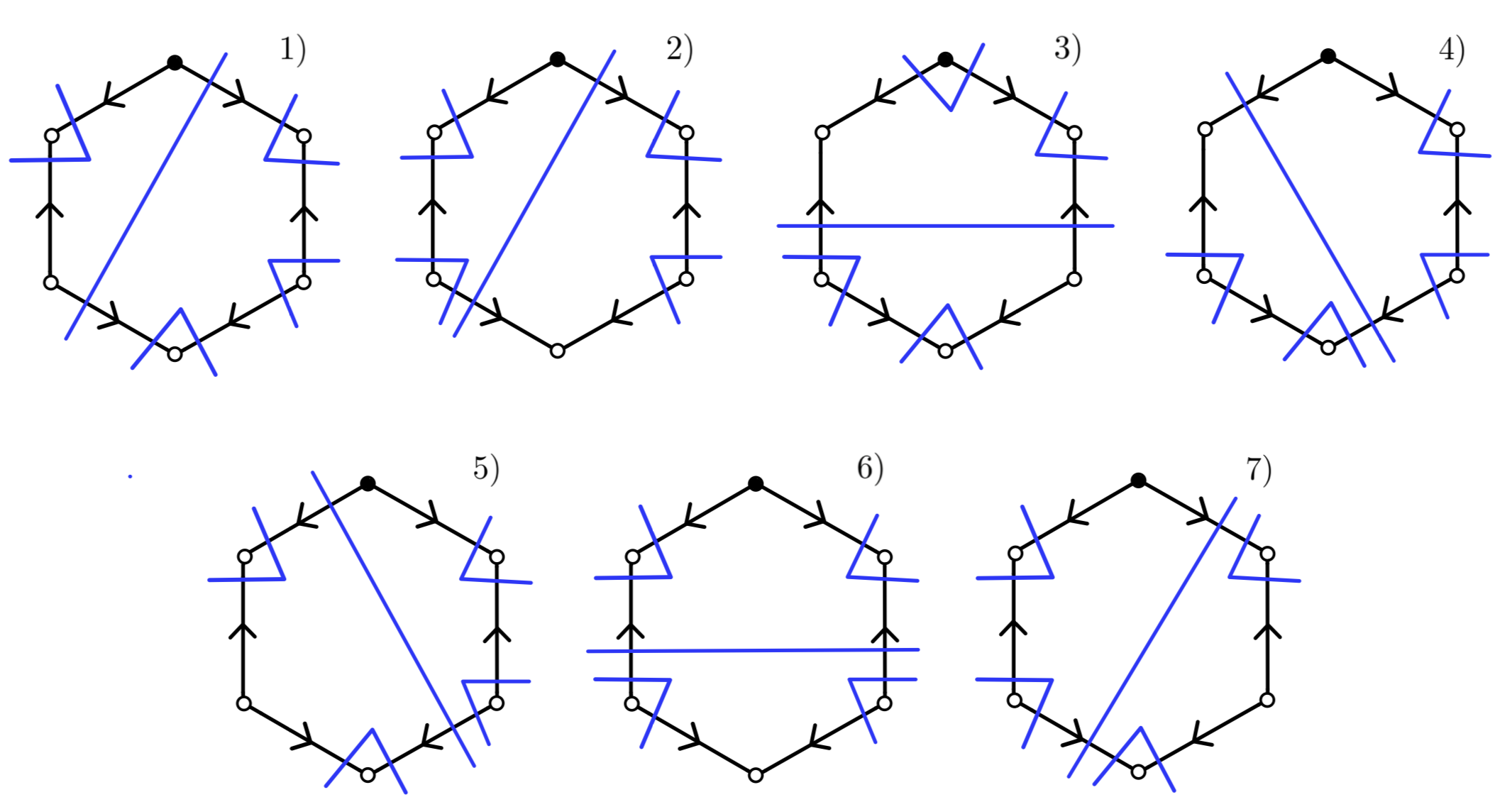}
\caption{Causal entangled thresholds allowed by selection rules 1-3, with 4 single-vertex and 1 three-vertex partitions. (Upper row) Generating configurations in $\Sigma_6|_{a}$. (Lower row) Configurations present in $\bar{\Sigma}^{\rm new}_6|_{a}$ after applying the fourth rule.}
\label{fig:Hexagon2}
\end{center}
\end{figure*}

So, we study in detail the configurations in $\left(O[{\cal R}_6](\bar{\Sigma}_6)\right)^{c}$. As we said, the dimension of $\left(O[{\cal R}_6](\bar{\Sigma}_6)\right)^{c}$ is 18, but it turns out that
\beq
\left(O[{\cal R}_6](\bar{\Sigma}_6)\right)^{c} \equiv O[{\cal R}_6](\Sigma_6|_{a}) \, ,
\eeq
with
\beqn
\nonumber \Sigma_6|_{a} &=& \{\{1, 2, 3, 5, 13\}, \{1, 2, 4, 5, 13\}, 
\\ && \{1, 3, 4, 6, 15\}, \{1, 2, 3, 4, 14\} \} \, ,
\eeqn
the generating set. Notice that all of them are composed by four single-vertex partitions (i.e. $\lambda_i$ with $i \in [1,6]$) and a tree-vertex partition (i.e. $\lambda_i$ with $i=\{13,14,15\}$), and this structure is obviously preserved by rotations. Therefore, let us isolate the configurations with these properties in $\bar{\Sigma}_6$ (i.e. those configurations already present in the standard causal representation obtained after applying the fourth selection rule):
\beqn
\nonumber \bar{\Sigma}_6^{\rm new} &=& \{\{1, 2, 3, 5, 14\}, \{1, 2, 4, 5, 15\}, 
\\ \nonumber && \{1, 3, 4, 5, 13\}, \{1, 3, 5, 6, 13\},
\\ &&  \{1, 3, 5, 6, 14\}\} \subset \bar{\Sigma}_6 \, ,
\eeqn
which can be further reduced by noticing that
\beq
\bar{\Sigma}_6^{\rm new} \equiv O[{\cal R}_6]\left(\bar{\Sigma}_6^{\rm new}|_{a}\right) \, ,
\eeq
where the generating set is given by
\beqn
\nonumber \bar{\Sigma}_6^{\rm new}|_{a} &=& \{\{1, 2, 3, 5, 14\}, \{1, 2, 4, 5, 15\}, 
\\ && \{1, 3, 4, 5, 13\}\} \, .
\eeqn
In Fig. \ref{fig:Hexagon2}, we show $\Sigma_6|_{a}$ (upper line) and $\bar{\Sigma}^{\rm new}_6|_{a}$ (lower lines) configurations; our aim is to find a transformation that relates both sets.

In order to do that, let us recall that $N$-gons are invariant under the action of the dihedral group $D_N$. This group is composed by $N$ rotations ($R_N$) and $N$ reflections ($S_N$). The rotations can be thought as the iterated application of ${\cal R}_N$, i.e. a clock-wise rotation with angle $\alpha = 2\pi/N$. The reflections are defined w.r.t. the symmetry axis of the $N$-gon.

As we show in Secs. \ref{sec:BoxExample} and \ref{sec:PentagonExample}, the elements of $\Sigma_N$ are related to those in $\bar{\Sigma}_N$ for $N=\{4,5\}$ through the application of $t \in R_N \subset D_N$. However, for $N \geq 6$ this is no longer true: the generating elements in Fig. \ref{fig:Hexagon2} form two disjoint sets under the application of all the transformations in $D_N$. This is because they have different topological properties, which appear when the diagram is dressed by the entangled causal thresholds. For instance, notice that the configuration 4 belonging to $\Sigma_6|_{a}$ (upper row in Fig. \ref{fig:Hexagon2}) has four adjacent $\lambda$'s and none of those in $\bar{\Sigma}^{\rm new}_6|_{a}$ fulfills that.

These properties will allow us to reformulate the graphic rules stated at the beginning, as we will show in the following.

\subsection{Delving into the selection rules}
\label{ssec:HexaAnalisis2}
Let us explore in more detail the differences between the first and the second row in Fig. \ref{fig:Hexagon2}. In first place, there is an interesting fact regarding configuration 1: the signature of the causal thresholds is not unique. If we do not impose the fourth rule, given a collection of causal thresholds, there are not enough constraints to fix the orientations of the inner lines. Then, we could have the term
\beq
\frac{1}{\lambda_1^+ \lambda_2^- \lambda_3^+ \lambda_{13}^+} \left(\frac{1}{\lambda_5^+} +\frac{1}{\lambda_5^-}\right) \, + (+ \longleftrightarrow -) \, ,
\eeq
that do not contribute to the singular structure of the scattering amplitude. In fact, it can be shown that this term is associated to a higher-order pole in $q_3=-q_4-p_3$ and its contribution should vanish. So, the fourth rule implies that given a covering of the graph with causal thresholds, there exists a unique DAG associated to it. 

To be more rigorous, let us consider an undirected graph $G=(V,E)$, the associated $\bar{\cal H}_E$ the $2^{(E-1)}$-dimensional Hilbert space of reduced orientations of the graph (i.e. fixing the direction of one edge to avoid a degeneration) and the set $\Sigma$ containing all the causal entangled threshold fulfilling rules 1-3. Then, we define $\Sigma' \subset \Sigma$ such that exist
\beq
{\cal F} : \Sigma' \to \bar{\cal H}_E \, ,
\eeq
and it is a surjective function. {In the case of MCG, the number of constrains imposed by rules 1-3 is high-enough to associate to each \emph{causal dressing of the diagram} (i.e. causal entangled thresholds) one and only one DAG. In the case of one-loop topologies, which are the extreme case of non-MCG graphs}, we need to impose:
\begin{enumerate}
\setcounter{enumi}{4}
    \item \emph{Non-degenerate orientations}: For each causal entangled threshold $\sigma \in \Sigma'$ one and only one DAG is associated.
\end{enumerate}
{It turns out that this selection criterion is equivalent to selection rule 4}. In the next section we will discuss new degeneration or symmetries derived from algebraic constrains, and establish the connection with the graph-theoretic approach.

\section{Relations among causal thresholds, kinematics and degeneration}
\label{sec:GeometryAlgebra}
{ In the following, we will establish a connection among causal thresholds and kinematical constrains. Whilst the formalism and most of the conclusions are still valid for multi-loop topologies, we focus in the one-loop case in order to ease the presentation of the results.

So,} let us take $\sigma \in \Sigma$ and randomly select $V-1$ internal momenta: without any loss of generality, we propose ${\cal I}=\{q_{1},\ldots,q_{V-1}\}$. Because of rule 1, $\sigma$ is a function of all the internal momenta (i.e. the elements of $I$), so we define a system of $V-1$ equations whose variables are the on-shell energies associated to the elements of ${\cal I}$. If
$\sigma = \{\lambda_{\sigma(1)},\ldots,\lambda_{\sigma(V-1)}\}$
with
\beq
\lambda_{\sigma(i)} = \sum_{j} a^j_{\sigma(i)} q^{(+)}_{j,0} + \sum_k b^k_{\sigma(i)} p_{k,0} \, ,
\eeq
then we can define the following $(V-1)\times(V-1)$ linear system $\lambda_{\sigma(i)}=0$, i.e.
\beqn
\nonumber \sum_{j=1}^{V-1} a^j_{\sigma(i)} q^{(+)}_{j,0} &=& - \sum_{j=V-1}^{|E|} a^j_{\sigma(i)} q^{(+)}_{j,0} 
\\ &-& \sum_k b^k_{\sigma(i)} p_{k,0} \, .
\eeqn
Let us call $A(\sigma)_{i j}=a_{\sigma(i)}^j$ the kinematic matrix of the entangled threshold $\sigma$. {For one-loop amplitudes with $N=\{4,5\}$, ${\rm Det}\left(A(\sigma)\right) \neq 0$ for every $\sigma \in \Sigma$ fulfilling selection rules 1-3. However, starting from the hexagon ($N=6$), this is no longer true.} 

If ${\rm Det}\left(A(\sigma)\right)=0$, then the system is not invertible and it has no unique solution. In particular, we have found that this implies that the momentum flow is degenerate or, equivalently, that two or more DAGs can be dressed with the same causal entangled threshold $\sigma$. An example of this situation is configuration 1 in Fig. \ref{fig:Hexagon2}. {At this point, we can establish a connection between the graph-theoretic and the algebraic criteria for the one-loop case}, rephrasing:
\begin{enumerate}
\setcounter{enumi}{4}
    \item \emph{Non-degenerate orientations} (algebra version): For each causal entangled threshold $\sigma \in \Sigma'$, ${\rm Det}(A(\sigma)) \neq 0$.
\end{enumerate}
Still, looking at the remaining missing configurations from Fig. \ref{fig:Hexagon2}, all of them fulfill this criterion and, in principle, could be used to write a causal representation in the LTD.

\subsection{Degeneration as graph mappings}
\label{ssec:DegMap}
Besides the relations derived for each individual causal entangled threshold $\sigma$, there are some underlying properties of the graph. {Again, in this subsection we focus exclusively on the one-loop case}. In first place, for the case of $N$-point one-loop diagrams, the number of possible causal thresholds $\lambda$ is given by
\beq
\#(\lambda) = \frac{V(V-1)}{2} \, ,
\eeq
with $V=N$ number of external legs and vertices. Then, each $\lambda$ is given by Eq. \ref{eq:LambdaDefinition}, so we can write 
\beqn
\lambda_j^{\pm} &=& \sum_{i \in I} (2 t^i_j - 1) \, q^{(+)}_{i,0} \pm \sum_{i \in O} (2 u^i_j - 1) p_{i,0} \, ,
\eeqn
where $t^i_{j} \equiv T$ and $u^i_{j}\equiv U$ are matrices in ${\mathbb Z}_2^{\#(\lambda) \times V}$. In particular, we can generalize the definitions given in Eqs. (\ref{eq:LambdasSimplesBOX}), (\ref{eq:LambdasSimplesPENTAGON}) and (\ref{eq:LambdasSimplesHEXAGON}), i.e.
\beqn
\lambda_i^\pm &=& q^{(+)}_{i,0} + q^{(+)}_{i+1,0} \pm p_{i,0} \ \, \,  i\in [1,V-1] \, ,
\\ \lambda_{N+1}^\pm &=& q^{(+)}_{1,0} + q^{(+)}_{3,0} \pm p_{12,0} \, ,
\eeqn
creating a $V \times V$ invertible linear system relating $\lambda$'s and on-shell energies, independently on the choice of causal thresholds signatures (i.e. $\lambda_i^+$ or $\lambda_i^-$ for each $i$). 

\begin{figure}[ht]
\begin{center}
\includegraphics[scale=0.13]{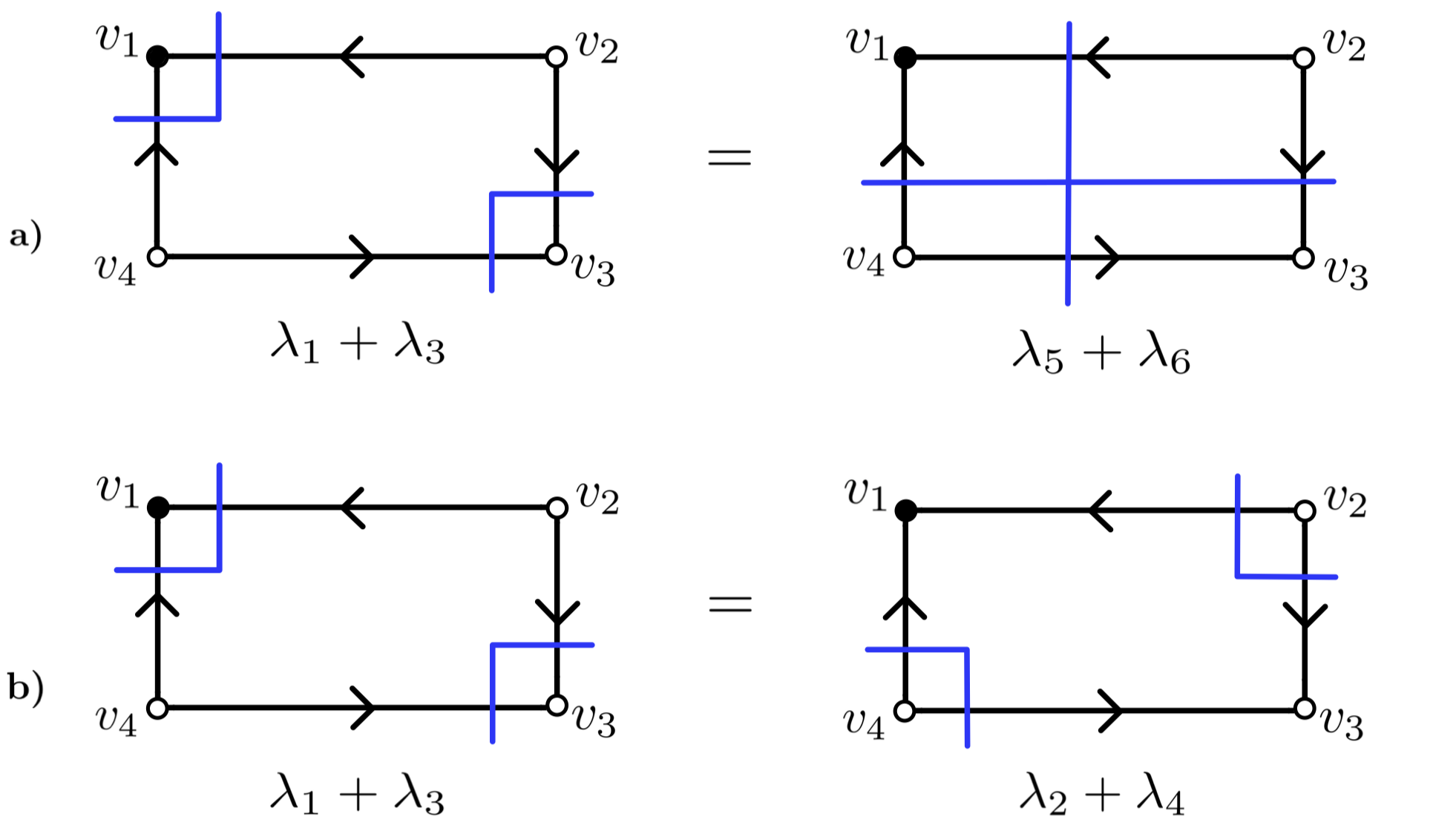}
\caption{graphic representation of the relation among causal thresholds for the box, given in Eqs. (\ref{eq:BoxRelal1l3l5l6}) and (\ref{eq:BoxRelal1l3l2l4}), in the upper and lower rows respectively.}
\label{fig:BoxRelations}
\end{center}
\end{figure}

For a random choice of signatures, the dependence on the external momenta will remain. So, let us first assume that all the external momenta are equal to 0, so  $\lambda_i^+=\lambda_i^-=\lambda_i$. Then, we can solve the system and write symbolically 
\beq
q^{(+)}_{j,0} = f(\lambda_1,\ldots,\lambda_{V-1},\lambda_{V}) \, ,
\eeq
with $j \in I=\{1,\ldots,V\}$ and $f$ a linear function. After that, we can write the remaining $\#(\lambda)-V$ causal thresholds in terms of $\{\lambda_1,\ldots,\lambda_{V-1},\lambda_{V}\}$, thus obtaining relations among the causal thresholds. As an example, for the box we have
\beqn
q^{(+)}_{1,0} &=& \frac{-\lambda_1+\lambda_2-\lambda_5}{2} \, ,
\\ q^{(+)}_{2,0} &=& \frac{-\lambda_1-\lambda_2+\lambda_5}{2} \, ,
\\ q^{(+)}_{3,0} &=& \frac{\lambda_1-\lambda_2-\lambda_5}{2} \, ,
\\ q^{(+)}_{4,0} &=& \frac{-\lambda_1+\lambda_2-2\lambda_3+\lambda_5}{2} \, ,
\eeqn
and we can rewrite the remaining causal thresholds, i.e. $\lambda_4$ and $\lambda_6$ as
\beqn
\lambda_4 &=& -\lambda_1+\lambda_2-\lambda_3 \, ,
\\ \lambda_6 &=& -\lambda_1-\lambda_3+\lambda_5 \, ,
\eeqn
which implies
\beqn
\lambda_1 +\lambda_3 &=& \lambda_5 +\lambda_6 \, ,
\label{eq:BoxRelal1l3l5l6}
\\  \lambda_1 +\lambda_3 &=& \lambda_2 +\lambda_4 \, .
\label{eq:BoxRelal1l3l2l4}
\eeqn
The next steps consist in translating these relations in terms of graphs, and then restoring the signature. For establishing the connection with graphs, let us consider Fig. \ref{fig:BoxRelations}. Now, we can write
\beqn
\nonumber && \frac{1}{\lambda_1 \lambda_5 \lambda_3} + \frac{1}{\lambda_1 \lambda_6 \lambda_3} = \frac{\lambda_5+\lambda_6}{\lambda_1\lambda_5\lambda_3\lambda_6} 
\\ &=& \frac{\lambda_1+\lambda_3}{\lambda_1\lambda_5\lambda_3\lambda_6}= \frac{1}{\lambda_3 \lambda_5 \lambda_6} + \frac{1}{\lambda_1 \lambda_5 \lambda_6} \, ,
\eeqn
and
\beqn
\nonumber && \frac{1}{\lambda_1 \lambda_2 \lambda_3} + \frac{1}{\lambda_1 \lambda_4 \lambda_3} = \frac{\lambda_2+\lambda_4}{\lambda_1\lambda_2\lambda_3\lambda_4} 
\\ &=& \frac{\lambda_1+\lambda_3}{\lambda_1\lambda_2\lambda_3\lambda_4}= \frac{1}{\lambda_1 \lambda_2 \lambda_4} + \frac{1}{\lambda_2 \lambda_3 \lambda_4} \, ,
\eeqn
where we applied Eqs. (\ref{eq:BoxRelal1l3l5l6}) and (\ref{eq:BoxRelal1l3l2l4}), respectively. 

Notice that, according to the graphic selection rules 1-3, $\lambda_5$ and $\lambda_6$ are incompatible and thus Eq. (\ref{eq:BoxRelal1l3l2l4}) is not relevant for the analysis of the symmetries of the causal representations. Also, we can restore the signatures simply by adding the external momenta in the definition of the causal thresholds and forcing their cancellation, obtaining
\beq
\lambda_1^\pm + \lambda_3^\pm = \lambda_2^\mp + \lambda_4^\mp \, .
\label{eq:BoxRelationVALID}
\eeq
The independence of these relations on the external momenta is guaranteed by the fact that $\sum p_i = 0$.

At this point, we observe that Eq. (\ref{eq:BoxRelal1l3l2l4}) is invariant under transformations in $D_4$. This implies that no new constrains, besides those arising from rotations or reflections of the graph, are introduced by the degeneration of causal thresholds.

For this simple example, we understand that there is an interplay among the number of symmetries of the graph (captured by $D_N$ for $N$-point one-loop functions) and the number of relations among $\lambda$'s. As we mention before, $D_N$ has $2 N$ elements, although in the case of one-loop diagrams we need only the $N=V$ rotations. On the other side, there are
\beq
\#(\lambda) - \#(I) = V(V-3)/2 \, ,
\eeq
relations among $\lambda$'s. For $N \leq 5$, there are more elements in $R_N$ than equations relating $\lambda$'s. Then, all the causal entangled thresholds can be connected by transformations in $R_N$ and this is why all the elements of $\Sigma_N$ are present in the symmetric causal representation from the LTD.

However, for $N \geq 6$, it turns out that $\#(\lambda) - \#(I) > N$ and this implies that some constraints cannot be linked through rotations: they live in disconnected sets. This explains why we cannot generate $\Sigma$ from simply symmetrizing $\bar{\Sigma}$ through rotations. We present a more detailed analysis of the hexagon case in App. \ref{app:Hexagon}.

\subsection{Minimally versus maximally connected graphs}
\label{ssec:Comparison}
{To conclude this discussion, let us explore how these ideas can be extended to study other topologies, and in particular, to the multi-loop case}. We start considering MCG, which are characterized by having the maximum possible number of edges. As the name suggests, every vertex in these graphs is interconnected, forming a fully connected structure. 

\begin{table}[ht]
\centering
\begin{tabular}{|c||c|c|}
\hline
 & \textbf{mCG (one-loop)} & \textbf{MCG (multiloop)} \\
\hline\hline
\(\#(\lambda)\) & \(V(V-1)/2\) & \(2^{V-1}-1\) \\
\hline
\(\#(I)\) & \(V\) & \(V(V-1)/2\) \\
\hline
Loops & \(1\) & \(V(V-2)/2\) \\
\hline
Relations & \(V(V-3)/2\) & \(2^{V-1}-1-V(V-2)/2\) \\
\hline
Symmetries & \(V\) & \(V!\) \\
\hline
\end{tabular}
\caption{Relevant characteristics to compare minimally connected (mCG) and maximally connected graphs (MCG).}
\label{tab:TablaResumen}
\end{table}

{In contrast to the one-loop case or minimally connected graph (mCG)}, the number of symmetries of this graph is maximal, being described by the symmetric group $S_N$ which has $N!$ elements. We compare in Tab. \ref{tab:TablaResumen} the properties of mCG and MCG. In particular, let us notice that
\beq
\#(S_N) > \#(\lambda) - \#(I) \, ,
\eeq
for every $N=V$ vertex topology. So, the number of constrains imposed by $S_V$ is much larger that the relations among $\lambda$'s derived from algebra. We conjecture that all the constraints derived from algebra can be derived from applying $s \in S_V$ to the underlying graph. Furthermore, in this case, the application of the graph-theoretic rules 1-3 is enough to recover the causal representation, thanks to the large number of conditions that each compatible causal entangled threshold must fulfill.

\section{Conclusions and outlook}
\label{sec:Conclusions}
In this article, we have explored the properties of the causal representations in the Loop-Tree Duality (LTD) from a geometric and algebraic perspective. We have observed in previous works \cite{Aguilera-Verdugo:2020kzc,Ramirez-Uribe:2020hes,Sborlini:2021owe} that this representation was not unique, and the main motivation of this work was shedding light on the origin of this degeneration.

By focusing on the one-loop case, we first noticed that the graph-theoretic selection rules from Ref. \cite{Sborlini:2021owe} manage to break the degeneration forcing more connections among the vertices. The application of rule 4 automatically removes some terms in order to recover exactly the same results obtained through purely algebraic methods \cite{TorresBobadilla:2021dkq,TorresBobadilla:2021ivx}. The conjecture presented in Ref. \cite{Sborlini:2016gbr} indicates that all the terms allowed by rules 1-3 correspond to compatible causal entangled thresholds (CET) which could be included within one causal representation. 

So, we compared the sets $\Sigma_N$ (rules 1-3) and $\bar{\Sigma}_N$ (rules 1-4). We used as a guiding principle the symmetries of the underlying graph, in this case $N$-gons, and we applied rotations to expand $\bar{\Sigma}_N$. In this way, we generated a symmetric causal representation under the action of $D_N$. Also, for $N\leq 5$, it turns out that the symmetric causal representation spanned over the full set $\Sigma_N$. 

For $N\geq6$, some elements of $\Sigma_N$ were not generated from the action of $D_N$ on $\bar{\Sigma}_N$. The reason behind this behavior was the existence of more relations among causal thresholds $\lambda$'s than transformations in $R_N \subset D_N$. We found a systematic way to derive such relations and prove the existence of disconnected sets.

Besides the one-loop case, we briefly analyzed the maximally connected graphs (MCG), i.e. the ones that contain the maximum number of loops for a given number of vertices. The fact that the symmetric group is larger than the number of relations among $\lambda$'s derived from algebraic equations supports that the symmetrization would be enough to generate all the elements in $\Sigma$. In any case, further studies are deferred for future works.

In conclusion, it is crucial to deepen our understanding of the symmetries underlying integrand-level representations of Feynman integrals. The causal representation is highly efficient because it contains only physical singularities, thereby avoiding terms that might introduce numerical instabilities. Moreover, the existence of nontrivial relations among combinations of causal entangled thresholds hints at an even more efficient representation and could serve as a powerful method for establishing connections among unsolved problems in geometry, algebra, and graph theory applied to Feynman integrals. In particular, the close connection among graph theory and combinatorial problems tractable with quantum algorithms could pave the road for the development of new and disruptive methods to efficiently calculate Feynman diagrams with quantum computers \cite{deLejarza:2022bwc,deLejarza:2023qxk,deLejarza:2024scm,deLejarza:2024pgk,Ramirez-Uribe:2021ubp,Clemente:2022nll}.

\section*{Acknowledgments}
We gratefully acknowledge G. Rodrigo and R. Hern\'andez-Pinto for reading the manuscript, R. Or\'us for inspiring discussions and Multiverse Computing for their generous hospitality during part of this project’s development. This work was supported by EU Horizon 2020 research and innovation programme STRONG-2020 project under Grant Agreement No. 824093 and H2020-MSCA-COFUND USAL4EXCELLENCE-PROOPI-391 project under Grant Agreement No 101034371.

\appendix
\section{More details about Causal Loop-Tree Duality}
\label{app:LTD}
{In order to give a self-contained presentation, we provide in this Appendix more details about LTD, its connection with causality and graph theory. Since the LTD-related literature has been evolving rapidly in the last few years, we try to summarize the main concepts and ideas here. Still, we recommend the interested reader to read Ref. \cite{Aguilera-Verdugo:2021nrn} for a more complete review of LTD and its applications.

Let us start with the main motivation behind LTD. In the context of HEP and SM-like theories, scattering amplitudes and Feynman integrals are defined in a four-dimensional Minkowski space-time. In the Feynman representation, a generic $N$-point $L$-loop scattering amplitude can be written as
\beqn
{\cal A}_N^{L} \approx \int_{\ell_1 \ldots \ell_L} \, {\cal N}(\{\ell_i\},\{p_i\}) \, \prod G_F(q_j) \, ,
\label{eq:GenGen}
\eeqn
with ${\cal N}$ a numerator depending on the loop momenta $\ell_i=(\ell_{i,0},\vec{\ell}_i)$, the external particle momenta $p_i=(p_{i,0},\vec{p}_i)$ and
\beq
\int_{\ell} = \mu^{4-d} \int \frac{d^{d}\ell}{(2\pi)^{d}}~,
\eeq
the integration measure in $d$-dimensions. The internal lines of the Feynman diagram are called \emph{edges} of the graph and connect \emph{vertices} which represent interacting points. Each edge corresponds to a propagating virtual state, being
\beq 
G_F(q_j) = \frac{1}{q_j^2-m_j^2+\imath 0} = \frac{1}{q_{j,0}-q_{j,0}^{(+)}} \times \frac{1}{q_{j,0}+q_{j,0}^{(+)}} \, ,
\label{eq:PropFeyn}
\eeq
the associated Feynman propagator in the scalar case and
\beq
q_{j,0}^{(+)} = \sqrt{|\vec{q}_j|^2+m_j-\imath 0} \, ,
\label{eq:q0plus}
\eeq
the on-shell positive energy for a particle with mass $m_j$ and four-momentum $q_j=(q_{j,0},\vec{q}_j)$. 

Notice that Eq. (\ref{eq:GenGen}) is fully general, even for fermions or vector-bosons propagating inside the diagram, since all the additional structure can be absorbed within ${\cal N}$. 

The key point of LTD is applying Cauchy residue theorem (CRT) to remove one degree of freedom per loop. In particular, we are interested in removing the energy component, in such a way that the remaining integral is defined on Euclidean space. This justifies the decomposition of Eq. (\ref{eq:PropFeyn}), since it makes explicit the pole structure of the propagators and thus, the pole structure of scattering amplitudes. As explained in Ref. \cite{Aguilera-Verdugo:2020nrp}, the application of CRT for multi-loop amplitudes involves the calculation of iterated residues selecting those poles with negative imaginary parts. For instance, let us consider a scalar two-loop sunrise diagram:
\beqn
\nonumber {\cal A}_2^2 &=& \int_{\ell_1,\ell_2} \frac{1}{q_1^2 - m_1^2 + \imath 0} \frac{1}{q_2^2 - m_2^2 + \imath 0} 
\\  &\times& \frac{1}{q_3^2 - m_3^2 + \imath 0} \, ,
\eeqn
with
\beqn
q_1 &=& \ell_1 \, ,
\\ q_2 &=& \ell_2 \, ,
\\ q_3 &=& \ell_1 +\ell_2 - p_1 \, ,
\eeqn
the momenta convention. Then, we remove the integration on $q_{1,0}=\ell_{1,0}$ by applying CRT closing the integration contour in the lower part of the complex plane \cite{Catani:2008xa}). The list of poles is
\beq
{\rm Poles}_{q_{1,0}} = \{\pm q_{1,0}^{(+)}, \pm q_{3,0}^{(+)}-q_{2,0}+p_{1,0}\} \, ,
\eeq
where we appreciate that only the on-shell energies $q_{1,0}^{(+)}$ and $q_{2,0}^{(+)}$ have non-vanishing imaginary part. So, we compute the residues for $q_{1,0} \to q_{1,0}^{(+)}$ and $q_{1,0} \to q_{3,0}^{(+)} -q_{2,0}+ p_{1,0}$, obtaining
\beqn
\nonumber {\cal A}_2^2 &=&  \int_{\vec{\ell}_1} \int_{\ell_2} \frac{1}{q_{2,0}^2-(q_{2,0}^{(+)})^2} \, \\ \nonumber &\times& \left[ \frac{1}{2 q_{1,0}^{(+)}((q_{1,0}^{(+)}+q_{2,0}-p_{1,0})^2-(q_{3,0}^{(+)})^2)} \right.
\\ &+& \left. \frac{1}{2 q_{3,0}^{(+)}((q_{3,0}^{(+)}-q_{2,0}+p_{1,0})^2-(q_{1,0}^{(+)})^2)} \right] \, .
\eeqn
After this, we need to compute the residue on $q_{2,0}$. The list of poles is
\beq
{\rm Poles}_{q_{2,0}} = \{\pm q_{2,0}^{(+)}, \pm q_{3,0}^{(+)}-q_{1,0}^{(+)}+p_{1,0}, q_{3,0}^{(+)}\pm q_{1,0}^{(+)}+p_{1,0}\} \, ,
\eeq
with
\beq
\int_{\vec{\ell}} = \imath \mu^{4-d} \int \frac{d^{d-1}\vec{\ell}}{(2\pi)^{d-1}}~.
\eeq
This time, we cannot unambiguously determine the sign of the imaginary part of $q_{3,0}^{(+)} - q_{1,0}^{(+)}$ so we can include their contribution multiplying by $\theta({\rm(q_{1,0}^{(+)} - q_{3,0}^{(+)})})$. As we rigorously prove in Ref. \cite{Aguilera-Verdugo:2020nrp}, these terms cancel and the final result only contains poles with well-defined imaginary part. In this case, we obtain
\beqn
A^2_2 &=&   \int_{\vec{\ell}_1,\vec{\ell}_2} \, 
\\ \nonumber && \left[ \frac{1}{4 q_{1,0}^{(+)} q_{2,0}^{(+)}} \frac{1}{\left(q_{1,0}^{(+)}+q_{2,0}^{(+)}+p_0\right)^2-q_{3,0}^{(+) 2}} \right.
\\ \nonumber &+& \left. \frac{1}{4 q_{1,0}^{(+)} q_{3,0}^{(+)}} \frac{1}{\left(q_{1,0}^{(+)}-q_{3,0}^{(+)}+p_0\right)^2-q_{2,0}^{(+) 2}}  \right.
\\ \nonumber &+&\left. \frac{1}{4 q_{2,0}^{(+)} q_{3,0}^{(+)}} \frac{1}{\left(q_{2,0}^{(+)}+q_{3,0}^{(+)}-p_0\right)^2-q_{1,0}^{(+) 2}} \right] \, .
\label{eq:Sunrisepre}
\eeqn
This is called the \emph{dual} representation of the sunrise diagram. Each term of the dual representation can be associated to a sunrise diagram with two on-shell edges \cite{Aguilera-Verdugo:2020kzc,Aguilera-Verdugo:2020nrp}, as depicted in Fig. \ref{fig:Sunrise} (left). These are called \emph{dual contributions} and each of them is mapped into a tree-level like object.

\begin{figure*}[ht]
\begin{center}
\includegraphics[scale=0.2]{Sunrisebuena.png}
\caption{{graphic analysis of the application of LTD to the two-loop sunrise diagram.
(a) Decomposition obtained after the iterated application of the Cauchy residue theorem. Each graph contains two cut or on-shell internal lines and can be opened into a tree. Moreover, each graph corresponds to one and only one term in Eq. (\ref{eq:Sunrisepre}).
(b) Causal representation of the sunrise, obtained by summing all the terms in Eq. (\ref{eq:Sunrisepre}). Each contribution corresponds to a threshold singularity of the graph.
(c) graphic definition of multi-edge, where all the edges connecting two given vertices are collapsed, and the corresponding positive on-shell energies are summed.}}
\label{fig:Sunrise}
\end{center}
\end{figure*}

However, if we add together these three terms, we obtain
\beqn
A^2_2 &=&  \int_{\vec{\ell}_1,\vec{\ell}_2} \frac{1}{8 q_{1,0}^{(+)} q_{2,0}^{(+)} q_{3,0}^{(+)}} \left[\frac{1}{\lambda^+}+\frac{1}{\lambda^-}\right] \, ,
\label{eq:CausalSun}
\eeqn
which is called a \emph{causal} representation and
\beq
\lambda^{(\pm)} = \sum_{i=1}^{3} q_{i,0}^{(+)} \pm p_{1,0} \, ,
\eeq
are the \emph{causal thresholds}. A graphic representation of Eq. (\ref{eq:CausalSun}) is given in Fig. \ref{fig:Sunrise} (center) where we appreciate the analogy with Cutkosky's cuts \cite{Cutkosky:1960sp}. The orientation of the internal lines indicates the energy flow: each term of Eq. (\ref{eq:CausalSun}) is mapped to a configuration with well-defined and fixed orientations. $\lambda^{(\pm)}=0$ are the only possible physical singularities of the integrand and they give rise to physical thresholds compatible with causality.

An interesting observation is that the functional form of Eq. (\ref{eq:CausalSun}) remains the same even when we consider a generalization of the sunrise with $L$-loops. This is pictorially shown in Fig. \ref{fig:Sunrise} (right) and motivates the definition of \emph{reduced Feynman graphs}, which are obtained from the original graphs by collapsing all the edges connecting the same vertices into a single \emph{multi-edge}. As we explained in Refs. \cite{Verdugo:2020kzh,Aguilera-Verdugo:2020nrp,Ramirez-Uribe:2020hes}, the reduced Feynman graph contains all the information required to describe the physical singularities of the corresponding scattering amplitude: \emph{Feynman graphs that share the same reduced form have the same causal representation}. 

\subsection*{Connection with graph theory}
Working at the level of reduced Feynman graphs, we are left with sets of vertices $V$ and multi-edges $E$ connecting them. By comparing several diagrams and their causal representations obtained from the calculation of iterated residues, we derived rules to describe them from the underlying graph $G=(V,E)$. 

In the first place, we observe that the set of possible causal thresholds $\{\lambda_i\}$ are in one-to-one correspondence to the connected binary partitions of $G$ \cite{Sborlini:2021owe}.   

Then, we notice that the number of vertices is related to the topological complexity or order of the diagram, $k=V-1$. In Eq. (\ref{eq:CausalSun}), we only have one causal threshold (i.e. one $\lambda$) per term. In the case of the box, described in Sec. \ref{sec:BoxExample}, we need to combine $k=4-1=3$ causal thresholds (i.e. three different $\lambda$'s) in each term. We can think about these combinations of thresholds as possible decompositions of the diagram that preserve the original physical singularities. Since not all the combinations of causal thresholds appear when adding together the dual terms, we introduce the concept of \emph{causal entangled thresholds} (CET) to refer to those present in causal representations. 

After studying causal representations of several multi-loop amplitudes starting from iterated residues, we conjecture Eq. (\ref{eq:causalRepresentation}). For simple topologies (i.e. with $k \leq 5$), Eq. (\ref{eq:causalRepresentation}) can be reconstructed from iterated residues using partial fractions or finite field techniques \cite{Ramirez-Uribe:2020hes}. However, as the order increases, the computational power required to compute the residues and then re-arrange the resulting expressions as in Eq. (\ref{eq:causalRepresentation}) becomes prohibitive. So, other strategies were explored. One possibility is using algebraic relations and conjecture a generic formula to describe the causal representation of maximally connected graphs (MCG), as discussed in Refs. \cite{TorresBobadilla:2021dkq,TorresBobadilla:2021ivx}. Another possibility, presented in Ref. \cite{Sborlini:2021owe}, relies on heuristic graph-theoretic selection rules for generating causal representations. Listed in Sec. \ref{ssec:GeometryReview}, these rules were obtained from the systematic study of several families of multi-loop topologies and the structure of their corresponding reduced graphs. 

In particular, it turns out that imposing a consistent orientation of the energy flow across the thresholds (rule 1) fixes the direction of the multi-edges of the reduced graph. Thus, causal representations are obtained from oriented or \emph{directed graphs} (DG). Strictly speaking, $G=(V,E)$ is an oriented graph if each multi-edge indicates the direction, i.e. given $e \in E$ then $e=\{v_i \to v_j\}$. Furthermore, selecting only those combinations of $k=V-1$ thresholds that do not cross each other (rule 3) implies that the graph must be compatible with a partition in such a way that the energy flow from one to the other is consistent. As heuristically shown in Refs. \cite{Ramirez-Uribe:2021ubp,Clemente:2022nll}, the combination of these two requirements implies that the oriented graph cannot contain cycles. This is the reason why each CET corresponds to a directed acyclic graph (DAG). In this way, causal representations can be bootstrapped by identifying all the DAG from the reduced Feynman graph and then \emph{dressing} them with the non-intersecting causal thresholds (rule 3).}

\section{Symmetric causal representation of the pentagon}
\label{app:Pentagon}
Here, we report the coefficients of the symmetric causal representation of the pentagon, present in Eq. (\ref{eq:SimetrizadaPentagon}). We have
\beqn
\nonumber C^{(1)}_5 &=& \frac{1}{\lambda^{-}_2 \lambda^{-}_5 \lambda^{+}_1 \lambda^{+}_3} + \frac{1}{\lambda^{-}_2 \lambda^{-}_5 \lambda^{+}_1 \lambda^{+}_4} +\frac{1}{\lambda^{-}_3 \lambda^{-}_5 \lambda^{+}_1 \lambda^{+}_4}
\\ &+&\frac{1}{\lambda^{-}_3 \lambda^{-}_5 \lambda^{+}_2 \lambda^{+}_4}+\frac{1}{\lambda^{-}_2 \lambda^{-}_4 \lambda^{+}_1 \lambda^{+}_3} \, ,
\eeqn
\beqn
\nonumber C^{(2)}_5 &=& \frac{1}{\lambda^{-}_3 \lambda^{-}_5 \lambda^{-}_8 \lambda^{+}_1}+\frac{1}{\lambda^{-}_2 \lambda^{-}_7 \lambda^{+}_1 \lambda^{+}_3}+\frac{1}{\lambda^{-}_4 \lambda^{-}_7 \lambda^{+}_1 \lambda^{+}_3}
\\ \nonumber &+& \frac{1}{\lambda^{-}_3 \lambda^{-}_{10} \lambda^{+}_2 \lambda^{+}_4}+\frac{1}{\lambda^{-}_3 \lambda^{-}_{10} \lambda^{+}_2 \lambda^{+}_5}+\frac{1}{\lambda^{-}_4 \lambda^{-}_7 \lambda^{+}_3 \lambda^{+}_5}
\\ \nonumber &+& \frac{1}{\lambda^{-}_3 \lambda^{+}_1 \lambda^{+}_4 \lambda^{+}_6}+\frac{1}{\lambda^{-}_5 \lambda^{+}_1 \lambda^{+}_4 \lambda^{+}_6}+\frac{1}{\lambda^{-}_3 \lambda^{+}_2 \lambda^{+}_4 \lambda^{+}_6}
\\ \nonumber &+& \frac{1}{\lambda^{-}_5 \lambda^{+}_2 \lambda^{+}_4 \lambda^{+}_6}+\frac{1}{\lambda^{-}_3 \lambda^{-}_5 \lambda^{+}_2 \lambda^{+}_7}+\frac{1}{\lambda^{-}_4 \lambda^{+}_2 \lambda^{+}_5 \lambda^{+}_8}
\\ \nonumber &+& \frac{1}{\lambda^{-}_4 \lambda^{+}_3 \lambda^{+}_5 \lambda^{+}_8}+\frac{1}{\lambda^{-}_2 \lambda^{+}_1 \lambda^{+}_3 \lambda^{+}_9}+\frac{1}{\lambda^{-}_5 \lambda^{+}_1 \lambda^{+}_3 \lambda^{+}_9}
\\ \nonumber &+& \frac{1}{\lambda^{-}_2 \lambda^{+}_1 \lambda^{+}_4 \lambda^{+}_9}+\frac{1}{\lambda^{-}_5 \lambda^{+}_1 \lambda^{+}_4 \lambda^{+}_9}+\frac{1}{\lambda^{-}_2 \lambda^{-}_4 \lambda^{+}_1 \lambda^{+}_{10}}
\\ &+& \frac{1}{\lambda^{-}_2 \lambda^{-}_5 \lambda^{+}_1 \lambda^{+}_{10}}+\frac{1}{\lambda^{-}_2 \lambda^{-}_5 \lambda^{-}_8 \lambda^{+}_1} \, ,
\eeqn
\beqn
\nonumber C^{(3)}_5 &=& \frac{1}{\lambda^{-}_2 \lambda^{-}_8 \lambda^{+}_1 \lambda^{+}_4}+\frac{1}{\lambda^{-}_3 \lambda^{-}_8 \lambda^{+}_1 \lambda^{+}_4}+\frac{1}{\lambda^{-}_3 \lambda^{-}_9 \lambda^{+}_2 \lambda^{+}_5}
\\ \nonumber &+& \frac{1}{\lambda^{-}_4 \lambda^{-}_9 \lambda^{+}_2 \lambda^{+}_5}+\frac{1}{\lambda^{-}_3 \lambda^{-}_5 \lambda^{+}_1 \lambda^{+}_6}+\frac{1}{\lambda^{-}_3 \lambda^{-}_5 \lambda^{+}_2 \lambda^{+}_6}
\\ \nonumber &+& \frac{1}{\lambda^{-}_5 \lambda^{+}_2 \lambda^{+}_4 \lambda^{+}_7}+\frac{1}{\lambda^{-}_4 \lambda^{+}_1 \lambda^{+}_3 \lambda^{+}_{10}}+\frac{1}{\lambda^{-}_5 \lambda^{+}_1 \lambda^{+}_3 \lambda^{+}_{10}}
\\ &+& \frac{1}{\lambda^{-}_2 \lambda^{-}_4 \lambda^{-}_7 \lambda^{+}_1} \, ,
\eeqn
and
\beqn
\nonumber C^{(0)}_5 &=& \frac{1}{\lambda^{-}_3 \lambda^{-}_7 \lambda^{-}_8 \lambda^{+}_1}+\frac{1}{\lambda^{-}_3 \lambda^{-}_7 \lambda^{-}_9 \lambda^{+}_1}+\frac{1}{\lambda^{-}_4 \lambda^{-}_7 \lambda^{-}_9 \lambda^{+}_1}
\\ \nonumber &+& \frac{1}{\lambda^{-}_3 \lambda^{-}_9 \lambda^{+}_1 \lambda^{+}_6}+\frac{1}{\lambda^{-}_4 \lambda^{-}_9 \lambda^{+}_1 \lambda^{+}_6}+\frac{1}{\lambda^{-}_3 \lambda^{-}_9 \lambda^{+}_2 \lambda^{+}_6}
\\ \nonumber &+&\frac{1}{\lambda^{-}_4 \lambda^{-}_9 \lambda^{+}_2 \lambda^{+}_6}+\frac{1}{\lambda^{-}_5 \lambda^{+}_2 \lambda^{+}_7 \lambda^{+}_8}+\frac{1}{\lambda^{-}_5 \lambda^{+}_3 \lambda^{+}_7 \lambda^{+}_8}
\\ \nonumber &+&\frac{1}{\lambda^{-}_5 \lambda^{+}_3 \lambda^{+}_7 \lambda^{+}_9}+\frac{1}{\lambda^{-}_5 \lambda^{+}_4 \lambda^{+}_7 \lambda^{+}_9}+\frac{1}{\lambda^{-}_4 \lambda^{+}_1 \lambda^{+}_6 \lambda^{+}_{10}}
\\ \nonumber &+&\frac{1}{\lambda^{-}_5 \lambda^{+}_1 \lambda^{+}_6 \lambda^{+}_{10}}+\frac{1}{\lambda^{-}_4 \lambda^{+}_2 \lambda^{+}_6 \lambda^{+}_{10}}+\frac{1}{\lambda^{-}_5 \lambda^{+}_2 \lambda^{+}_6 \lambda^{+}_{10}}
\\ \nonumber &+&\frac{1}{\lambda^{-}_4 \lambda^{+}_2 \lambda^{+}_8 \lambda^{+}_{10}}+\frac{1}{\lambda^{-}_5 \lambda^{+}_2 \lambda^{+}_8 \lambda^{+}_{10}}+\frac{1}{\lambda^{-}_4 \lambda^{+}_3 \lambda^{+}_8 \lambda^{+}_{10}}
\\  &+&\frac{1}{\lambda^{-}_5 \lambda^{+}_3 \lambda^{+}_8 \lambda^{+}_{10}}+\frac{1}{\lambda^{-}_2 \lambda^{-}_7 \lambda^{-}_8 \lambda^{+}_1}\, .
\eeqn
All the causal thresholds' definitions, the standard causal representation and the dual representation obtained from explicit residue calculations are available in Zenodo \cite{ZENODO}. In this repository, we include the results for $N$-point one-loop functions up to $N=9$.

\begin{figure*}[ht]
\begin{center}
\includegraphics[scale=0.34]{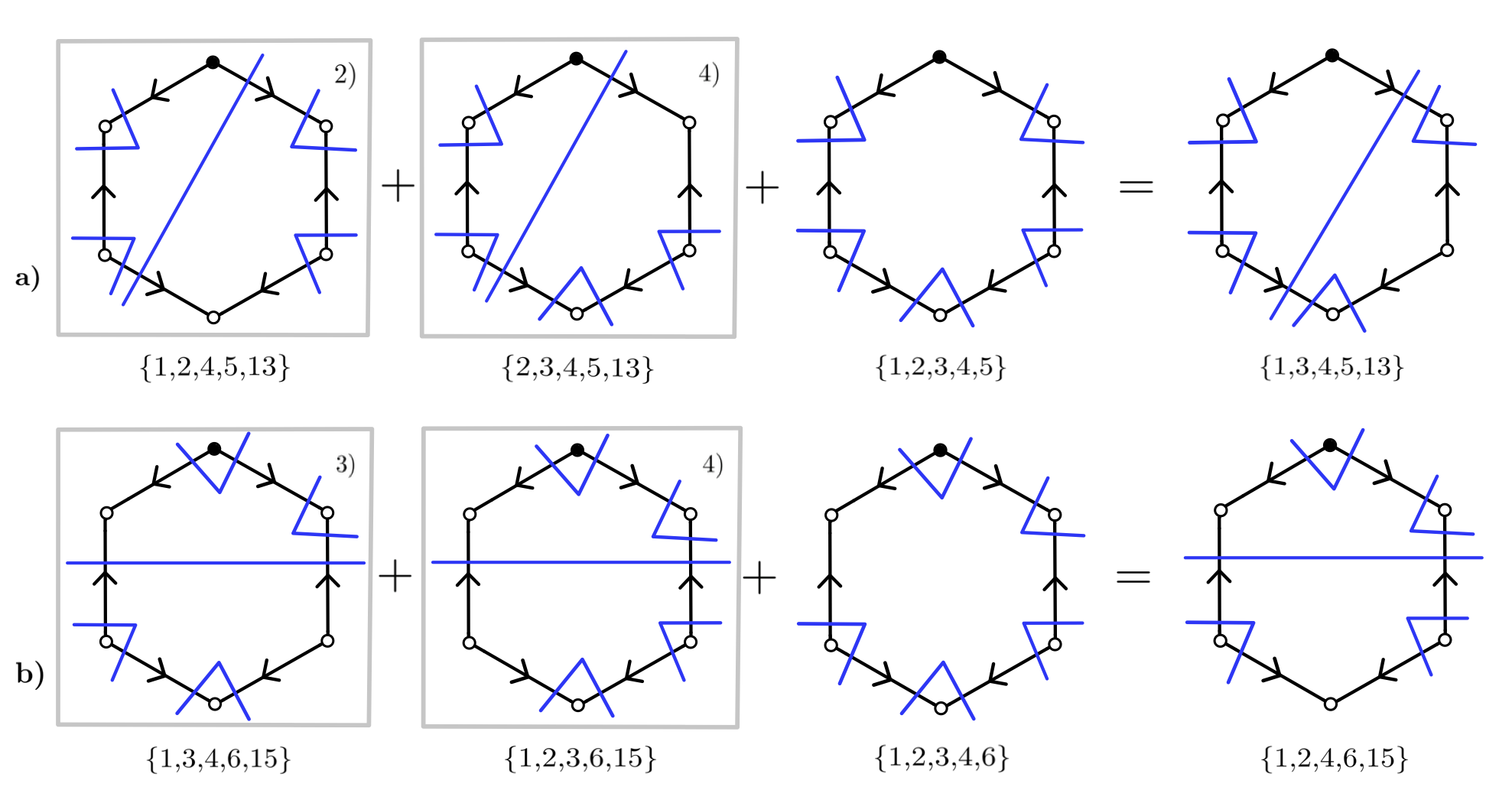}
\caption{Implementation of Eq. (\ref{eq:HexB4}) to configurations absent in the symmetrized causal representation for the hexagon, described in the upper row of Fig. \ref{fig:Hexagon2}. (Upper row) Expansion of configuration (2), i.e. $\{1,2,4,5,13\}$. (Lower row) Expansion of configuration (4), i.e. $\{1,3,4,6,15\}$. The elements inside boxes are part of $O[{\cal R}_6](\bar{\Sigma}_6^{\rm new}|_a)$, whilst the others are already present in the symmetrized causal representation.}
\label{fig:HexagonLAST}
\end{center}
\end{figure*}

\section{Analysis of the degeneration of causal entangled thresholds in the hexagon}
\label{app:Hexagon}
By following the procedure detailed in Sec. \ref{ssec:DegMap}, we first create the system of equations to write $\{q^{(+)}_{i,0}\}_{i=1,\ldots,6}$ as functions of $\{\lambda_1,\lambda_2,\lambda_3,\lambda_4,\lambda_5,\lambda_7\}$ neglecting the external momenta. We obtain:
\beqn
q^{(+)}_{1,0} &=&(-\lambda_{1}+\lambda_{2}-\lambda_{7})/2 \, ,
\\ q^{(+)}_{2,0} &=& (-\lambda_{1}-\lambda_{2}+\lambda_{7})/2\, ,
\\ q^{(+)}_{3,0} &=& (\lambda_{1}-\lambda_{2}-\lambda_{7})/2\, ,
\\ q^{(+)}_{4,0} &=& (-\lambda_{1}+\lambda_{2}-2 \lambda_{3}+\lambda_{7}) / 2\, ,
\\ q^{(+)}_{5,0} &=&  (\lambda_{1}-\lambda_{2}+2 \lambda_{3}-2 \lambda_{4}-\lambda_{7})/2\, ,
\\ q^{(+)}_{6,0} &=& (-\lambda_{1}+\lambda_{2}-2 \lambda_{3}+2 \lambda_{4}-2 \lambda_{5}+\lambda_{7}) / 2\, .
\eeqn
Then, we write the remaining causal thresholds as:  
\beqn
\lambda_{6}&=&-\lambda_{1}+\lambda_{2}-\lambda_{3}+\lambda_{4}-\lambda_{5}\, ,
\label{eql6}
\\ \lambda_{8}&=&-\lambda_{1}-\lambda_{3}+\lambda_{4}-\lambda_{5}+\lambda_{7}\, ,
\label{eql8}
\\ \lambda_{9}&=&-\lambda_{1}-\lambda_{3}+\lambda_{7} \, ,
\label{eql9}
\\ \lambda_{10}&=&\lambda_{1}-\lambda_{2}+\lambda_{3}-\lambda_{4}-\lambda_{7}\, 
\label{eql10}
\\ \lambda_{11}&=&-\lambda_{1}+\lambda_{2}-2 \lambda_{3}+\lambda_{4}-\lambda_{5}+\lambda_{7}\, ,
\label{eql11}
\\ \lambda_{12}&=&\lambda_{3}-\lambda_{4}-\lambda_{7}\, ,
\label{eql12}
\\ \lambda_{13}&=&-\lambda_{1}+\lambda_{2}-\lambda_{3}\, ,
\label{eql13}
\\ \lambda_{14}&=&-\lambda_{3}+\lambda_{4}-\lambda_{5}\, ,
\label{eql14}
\\ \lambda_{15}&=&-\lambda_{2}+\lambda_{3}-\lambda_{4}\, .
\label{eql15}
\eeqn
The next step consists in recombining appropriately Eqs. (\ref{eql6})-(\ref{eql15}) in such a way that we can map them into combinations of graphs. In particular, we look for equations of the generic form:
\beq
\lambda_{l_1} + \lambda_{l_2} + \ldots = \lambda_{r_1} + \lambda_{r_2} + \ldots \, .
\eeq
After some rearranging, we obtain three different sets of relations. The first one
\beqn
\lambda_2+ \lambda_4 &=& \lambda_1 +\lambda_3 + \lambda_5 +\lambda_6 \, ,
\\ \lambda_1 + \lambda_3 &=& \lambda_2 +\lambda_4 + \lambda_7 +\lambda_{10} \, ,
\\ \lambda_4 + \lambda_7 &=& \lambda_1 +\lambda_3 + \lambda_5 +\lambda_8 \, ,
\eeqn
relates 2 terms in the l.h.s. with 4 in the r.h.s. The second
\beqn
\lambda_7 &=& \lambda_1 +\lambda_3 + \lambda_9 \, ,
\\ \lambda_3 &=& \lambda_4 +\lambda_7 + \lambda_{12} \, ,
\\ \lambda_2 &=& \lambda_1 +\lambda_3 + \lambda_{13} \, ,
\\ \lambda_3 &=& \lambda_2 +\lambda_4 + \lambda_{15} \, ,
\\ \lambda_4 &=& \lambda_3 +\lambda_5 + \lambda_{14} \, ,
\eeqn
relates 1 term in the l.h.s with 3 in the r.h.s. Finally, we find
\beq
\lambda_3+\lambda_5+\lambda_{10}+\lambda_{11} = 0 \, ,
\eeq
which indicates that the combination of 4 different causal thresholds must vanish because of kinematical constraints. It is crucial to notice that these three sets cannot be related applying any symmetry transformation in $D_6$, simply because they involve different number of causal thresholds. Also, relying on the graph-theoretic selection rule 1-3, some of the equations involve incompatible causal thresholds. Imposing these rules, restoring the signatures and removing the degeneration due to rotations, we obtain:
\beqn
\lambda^+_2+ \lambda^+_4 &=& \lambda^-_1 +\lambda^-_3 + \lambda^-_5 +\lambda^-_6 \, ,
\label{eq:HexB1}
\\ \lambda^+_1 + \lambda^+_3 &=& \lambda^-_2 +\lambda^-_4 + \lambda^+_7 +\lambda^+_{10} \, ,
\label{eq:HexB2}
\\ \lambda^+_3 &=& \lambda^-_4 +\lambda^-_7 + \lambda^-_{12} \, ,
\label{eq:HexB3}
\\ \lambda^+_2 &=& \lambda^-_1 +\lambda^-_3 + \lambda^+_{13} \, .
\label{eq:HexB4}
\eeqn
Translating these equations into causal entangled thresholds, we proceed as previously done for the box. We combine the thresholds in the l.h.s. (index $l_i$) with those in the r.h.s. (index $r_i$) according to 
\beqn
\nonumber && \{r_1,r_2,r_3,r_4,l_1\} + \{r_1,r_2,r_3,r_4,l_2\}  
\\ \nonumber &=& \{l_1,l_2,r_1,r_2,r_3\}+ \{l_1,l_2,r_1,r_2,r_4\} 
\\ &+&  \{l_1,l_2,r_1,r_3,r_4\}+ \{l_1,l_2,r_2,r_3,r_4\}  \, ,
\eeqn
for Eqs. (\ref{eq:HexB1})-(\ref{eq:HexB2}) and 
\beqn
\nonumber \{r_1,r_2,r_3,X\} &=& \{l_1,r_1,r_2,X\} + \{l_1,r_1,r_3,X\}
\\ &+&  \{l_1,r_2,r_3,X\}  \, ,
\eeqn
for Eqs. (\ref{eq:HexB3})-(\ref{eq:HexB4}), where $X=\lambda_{a_1} \lambda_{a_2}$ involves causal thresholds not present in the l.h.s neither in the r.h.s. of the previously mentioned equations.

Now, let us show that the elements absent in the symmetrized causal representation of the hexagon can be generated using Eqs. (\ref{eq:HexB1})-(\ref{eq:HexB4}). We can corroborate that none of the equations is applicable to (1), which is also in agreement with the fact that this configuration is forbidden by rule 5. The remaining configurations are decomposed using Eq. (\ref{eq:HexB3}) as graphically shown in Fig. \ref{fig:HexagonLAST}. Restoring the signatures, we can write (upper row in Fig. \ref{fig:HexagonLAST}):
\beqn
\nonumber && \frac{1}{\lambda^+_1 \lambda^-_2 \lambda^-_4 \lambda^+_5 \lambda^+_{13}} + \frac{1}{\lambda^-_2 \lambda^+_3 \lambda^-_4 \lambda^+_5 \lambda^+_{13}} 
\\ &+& \frac{1}{\lambda^+_1 \lambda^-_2 \lambda^+_3 \lambda^+_5 \lambda^+_{13}} = \frac{1}{\lambda^+_1 \lambda^+_3 \lambda^-_4 \lambda^+_5 \lambda^+_{13}} \, ,
\eeqn
and (lower row in Fig. \ref{fig:HexagonLAST}):
\beqn
\nonumber && \frac{1}{\lambda^+_1 \lambda^+_3 \lambda^-_4 \lambda^-_6 \lambda^-_{15}} + \frac{1}{\lambda^+_1 \lambda^-_2 \lambda^+_3 \lambda^-_6 \lambda^-_{15}} 
\\ &+& \frac{1}{\lambda^+_1 \lambda^-_2 \lambda^+_3 \lambda^-_4 \lambda^-_6} = \frac{1}{\lambda^+_1 \lambda^-_2 \lambda^-_4 \lambda^-_6 \lambda^-_{15}} \, ,
\eeqn
for configurations (2) and (3) from Fig. \ref{fig:Hexagon2}, respectively. Notice that both expansions involve a configuration that can be connected to (4) using rotations. The remaining elements in these equations are already present in the symmetrized causal representation.

Thus, we have explicitly shown that there are transformations beyond $D_N$ that relate different causal entangled thresholds, and could be further exploited to achieve a more compact causal representation. We defer the study on this topic for future publications.


%

\end{document}